\documentclass[12pt, letterpaper]{JHEP3}

\pdfoutput=1

\usepackage{amssymb, amsmath, amsopn, amsthm, dsfont}
\usepackage{epsfig}
\usepackage{psfrag}
\usepackage{subfigure}

\newcommand{\eref}[1]{(\ref{#1})}

\newcommand{\cref}[1]{Chapter~\ref{#1}}
\newcommand{\beq}{\begin{equation}}
\newcommand{\eeq}{\end{equation}}
\newcommand{\ba}{\begin{array}}
\newcommand{\ea}{\end{array}}
\newcommand{\bcenter}{\begin{center}}
\newcommand{\ecenter}{\end{center}}

\def\IB{\relax\hbox{$\inbar\kern-.3em{\rm B}$}}
\def\IC{\relax\hbox{$\inbar\kern-.3em{\rm C}$}}
\def\ID{\relax\hbox{$\inbar\kern-.3em{\rm D}$}}
\def\IE{\relax\hbox{$\inbar\kern-.3em{\rm E}$}}
\def\IF{\relax\hbox{$\inbar\kern-.3em{\rm F}$}}
\def\IG{\relax\hbox{$\inbar\kern-.3em{\rm G}$}}
\def\IGa{\relax\hbox{${\rm I}\kern-.18em\Gamma$}}
\def\IH{\relax{\rm I\kern-.18em H}}
\def\IK{\relax{\rm I\kern-.18em K}}
\def\IL{\relax{\rm I\kern-.18em L}}
\def\IP{\relax{\rm I\kern-.18em P}}
\def\IR{\relax{\rm I\kern-.18em R}}
\def\IZ{\relax\ifmmode\mathchoice
{\hbox{\cmss Z\kern-.4em Z}}{\hbox{\cmss Z\kern-.4em Z}}
{\lower.9pt\hbox{\cmsss Z\kern-.4em Z}}
{\lower1.2pt\hbox{\cmsss Z\kern-.4em Z}}\else{\cmss Z\kern-.4em Z}\fi}
\def\II{\relax{\rm I\kern-.18em I}}

\def\sCC{{\kern 0.27em\vrule height1.45ex width0.03em depth0em
          \kern-0.30em\rm C}}
\def\C{{\mathchoice
  {\sCC}
  {\sCC}
  {\kern 0.225em \vrule height1.05ex width0.025em depth0em \kern-0.25em \rm C}
  {\kern 0.180em \vrule height0.78ex width0.02em depth0em \kern-0.2em \rm C}
        }}
\def\sHH{{\rm I\kern-.16em{}H}}
\def\H{{\mathchoice
  {\sHH}
  {\sHH}
  {\rm I\kern-.13em{}H}
  {\rm I\kern-.13em{}H} }}
\def\sNN{{\rm I\kern-.16em{}N}}
\def\N{{\mathchoice
  {\sNN}
  {\sNN}
  {\rm I\kern-.12em{}N}
  {\rm I\kern-.10em{}N} }}
\def\sPP{{\rm I\kern-.16em{}P}}
\def\P{{\mathchoice
  {\sPP}
  {\sPP}
  {\rm I\kern-.12em{}P}
  {\rm I\kern-.10em{}P} }}
\def\sQQ{{\kern 0.27em \vrule height1.45ex width0.03em depth0em
          \kern-0.30em \rm Q}}
\def\Q{{\mathchoice
        {\sQQ}
        {\sQQ}
  {\kern 0.225em \vrule height1.05ex width0.025em depth0em \kern-0.25em \rm Q}
  {\kern 0.180em \vrule height0.78ex width0.020em depth0em \kern-0.20em \rm Q}
        }}
\def\sRR{{\rm I\kern-0.16em{}R}}
\def\R{{\mathchoice
  {\sRR}
  {\sRR}
  {\rm I\kern-0.12em{}R}
  {\rm I\kern-0.10em{}R} }}
\def\sZZ{{\rm Z\kern-0.32em{}Z}}
\def\Z{{\mathchoice
  {\sZZ}
  {\sZZ} 
  {\rm Z\kern-0.3em{}Z}     
  {\rm Z\kern-0.25em{}Z} }}  
\def\ZZZ{{\rm Z\kern-0.24em{}Z}}
\def\sII{{\rm I\kern-0.16em{}I}}
\def\I{{\mathchoice
  {\sII}
  {\sII}
  {\rm I\kern-0.12em{}I}
  {\rm I\kern-0.10em{}I} }}

\def\Tr{{\rm Tr}}

\def\inbar{\,\vrule height1.5ex width.4pt depth0pt}
\font\cmss=cmss10 \font\cmsss=cmss10 at 7pt

\def\smiley{\hbox{\large$\bigcirc$\hspace{-0.80em}\raise.2ex
\hbox{$\cdot\cdot$}\kern-.61em\lower.2ex\hbox{\scriptsize$\smile$}}\ }
\def\frowny{\hbox{\large$\bigcirc$\hspace{-0.80em}\raise.2ex
\hbox{$\cdot\cdot$}\kern-.635em\lower.2ex\hbox{\scriptsize$\frown$}}\ }

\def\I{{\rlap{1} \hskip 1.6pt \hbox{1}}}

\makeatletter
\let\hangafter\@hangfrom
\makeatother

\newcommand{\calF}{{\cal F}}

\newcommand{\onebb}{\mathds{1}}
\newcommand{\Cbb}{\mathds{C}}

\newcommand{\Zbb}{\mathds{Z}}

\title{N-ification of Forces:\\ {\Large A Holographic Perspective on D-brane Model Building}}

\author{Sebasti\'an Franco$^1$, Diego Rodr\'iguez-G\'omez$^{1,2}$ and Herman Verlinde$^1$

\\
~\\
$^1$Joseph Henry Laboratories, Princeton University \\
Princeton, NJ 08544, USA \\ \vspace{0.3cm}
$^2$ Center for Research in String Theory, Queen Mary University of London \\
Mile End Road, London, E1 4NS, UK \\ \vspace{0.3cm}

\email{sfranco,drodrigu,verlinde@Princeton.EDU}\\
}

\abstract{We study geometric aspects of extensions of the supersymmetric standard model that exhibit a periodic
duality cascade. Via the holographic correspondence, the growth of the gauge group rank towards the UV is interpreted 
as a gradual decompactification transition. We show that this class of models 
typically 
develop a duality wall in the UV, and present an efficient method for estimating the hierarchy between
the on-set of the cascade and the formation of the wall.
As an illustrative example, we study the model introduced by Cascales, Saad and Uranga in [14],
which has an known geometric realization in terms of D-branes on an SPP$/\Zbb_3$ singularity. 
}

\preprint{PUPT-2263 \\ QMUL-PH-08-07}

\newcommand{\half}{\frac{1}{2}}

\def\be{\begin{equation}}
\def\ee{\end{equation}}
\def\bea{\begin{eqnarray}}
\def\eea{\end{eqnarray}}

\renewcommand{\footnotesize}{\small}
\begin{document}

\addtolength{\baselineskip}{.4mm}
\addtolength{\parskip}{.5mm}

\addtolength{\abovedisplayskip}{1mm}
\addtolength{\belowdisplayskip}{1mm}

\tableofcontents

\pagebreak

\section{Introduction}

\label{section_introduction}

Supersymmetric Grand Unification still stands out as the most compelling
paradigm for beyond the Standard Model physics -- it is elegant,  explains
the value of $\sin \theta_W$, and can be naturally embedded in string theory \cite{Dimopoulos:1981yj}.\footnote{See \cite{Raby:2006sk} for a review of the current status.} Several  developments within string theory, however, have motivated the
exploration of an array of alternative possibilities.  In particular, the
discoveries of D-branes \cite{Polchinski:1995mt}, warped flux compactifications, and the string
landscape \cite{Douglas:2006es} have led to the recognition that supersymmetric GUTs form only
a very small subset among many conceivable string realizations of the
Standard Model. 

A common feature of all string (motivated) phenomenological scenarios is that  
4-d effective field theory breaks down at some high energy scale, above which extra dimensional or string physics takes over. 
 Traditionally, the decompactification scale is taken to be close to the GUT scale, but in many popular scenarios, it is assumed to be much lower.  
 From the 4-d vantage point, the decompactification transition manifests itself
via a sudden escalation in the number of field theoretic degrees of freedom.
Any well thought out phenomenological scenario based on extra dimensions must therefore
supply  a cogent description of the dynamics among all these extra dimensional modes, and of their interactions  with Standard Model particles. A convincing way of specifying these dynamical rules is by
finding a natural realization of the given extra dimensional scenario within string theory. 

A popular class of low scale scenarios postulates that the Standard Model degrees of freedom
reside inside a 5-d 
warped geometry, either localized on branes or more smoothly distributed along the extra dimension \cite{Randall:1999ee}. 
An attractive aspect of this type of models is that the warped extra dimensional physics 
can be viewed through  the prism of  the AdS/CFT correspondence \cite{Maldacena:1997re,Gubser:1998bc,Witten:1998qj}, and reformulated in pure 4-d 
language in terms of the dual strongly coupled large $N$ gauge theory.  
The AdS/CFT dictionary comes with precise rules, which, if followed correctly, 
supply a consistent string embedding of the given warped scenario.

Via the holographic perspective, the decompactification transition 
is recast as a sudden expansion of the total effective
rank $N$ (or number of  `colors') of the 4-d gauge theory.
This precipitous increase in the number of colors may arise via symmetry restoration or deconfinement of a large $N$ gauge group. 
Models with large warped extra dimensions are
extreme representatives of a continuous family of beyond-the-Standard-Model physics scenarios
in which the total rank increases with energy scale.

A natural way for the number of colors to run with scale is via a duality cascade \cite{Klebanov:2000hb,Strassler:2005qs}.
In the context of renormalization group flows, a change in effective  rank 
arises when an asymptotically free gauge theory flows to strong coupling in the IR, where its low 
energy physics allows a weakly coupled description in terms of a dual theory with fewer colors.
A duality cascade is the general phenomenon whereby a quiver gauge theory undergoes a 
sequence of Seiberg dualities. Cascading RG flows are of particular interest because they occur for many field theories
with known supergravity duals \cite{Klebanov:2000hb,Franco:2004jz,
Herzog:2004tr}. The classic
example 
is the Klebanov-Strassler (KS) gauge theory, the holographic dual to the fluxed deformed conifold \cite{Klebanov:2000hb}.

The existence of a holographic dual  is directly linked to
the realization of the gauge theory via open strings attached to a stack of D-branes.
Starting with just a few D-branes in the IR, near the bottom
of the warped throat, the cascade will drive the effective number of branes to 
increase towards the UV, until it is large enough to be captured by a dual geometry.

As  noted by several authors \cite{Strassler,
Cascales:2005rj,Heckman:2007zp},  the existence of duality cascades  suggests an interesting
new avenue for model building. In the IR we observe the Standard Model, which has a finite rank gauge group. 
In the UV, on the other hand, string theory motivates the possibility that physics becomes 
higher dimensional and may admit, over some energy range, a dual description in terms of a large $N$ 4-d gauge theory. It is then natural to consider the general class of scenarios in which the Standard Model in the IR and the large $N$ theory in the UV are smoothly connected.
We will refer to this class of new physics models as `N-ification scenarios'.

\bigskip

\subsection*{N-ification: Decompactification via a Duality Cascade}

As indicated in Fig.~1, N-ification scenarios  in a 
sense interpolate between Grand Unified models, that assume a relatively minimal field content
below the GUT scale, and large/warped extra dimensional scenarios \cite{ArkaniHamed:1998rs,Randall:1999ee}, that postulate the
appearance of many new degrees of freedom at a relatively low scale. In N-ification models, the
the decompactification process starts at a low scale, but takes place gradually.

There are many possible ways in which the MSSM may connect to a 
cascading gauge theory in the UV. Starting in the IR, 
the first step upwards in the cascade introduces extra gauge and matter degrees of freedom. 
These extra particles are invisible below some scale $\Lambda_c$, 
at which they become confined or acquire a mass via symmetry breaking.
For any realistic scenario,  the experimental limits on compositeness 
or existence of exotic matter dictate a lower bound for the critical scale $\Lambda_c$
in the multi-TeV range.

Given the first step upwards in the cascade, and assuming that no other 
new gauge groups appear at some higher scale, the UV physics in principle follows from integrating the RG flow  upwards. Whenever one of the gauge groups reaches strong coupling, one can use Seiberg duality to switch to
a weakly coupled electric formulation.  Without extra guidance
from string theory or some other UV-motivated principle, however, the bottom-up 
perspective typically leads to irregular duality sequences, that involve rapidly growing 
ranks and numbers of generations. A more
attractive class of models are the periodic scenarios, in which the structure of the quiver diagram 
and the number of generations  remain constant  throughout the cascade, and only the 
ranks of the gauge groups change. Besides being
more appealing as field theories, such regularly cascading theories are also more likely to
allow a realization as the world-volume theory of D-branes at some suitable geometric
singularity.

\begin{figure}[t]
\begin{center}
 \epsfig{figure=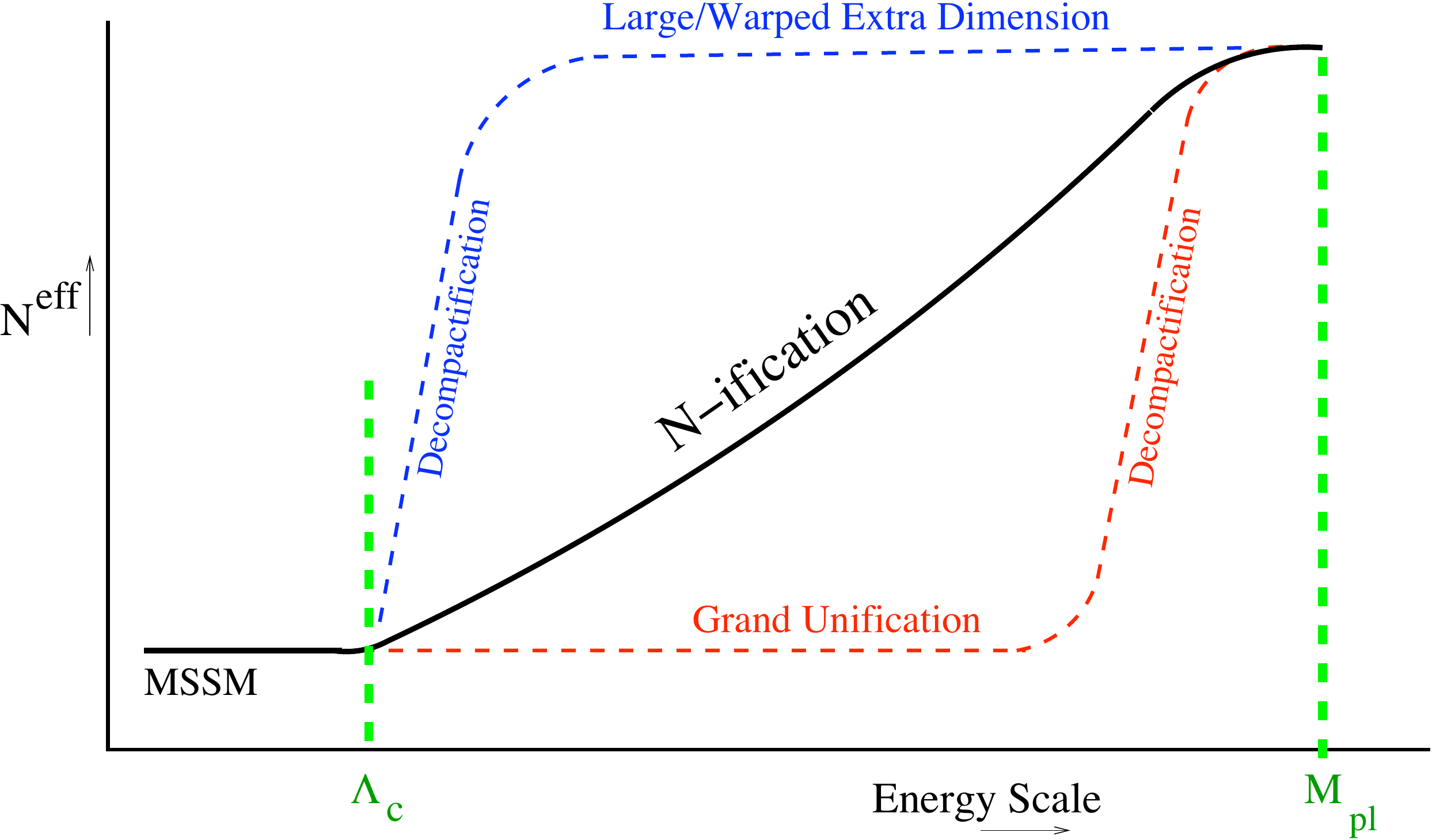,scale=.5}
\end{center}
  \caption{The number of degrees of freedom as a function of energy scale for three
  scenarios. In Grand Unified models, the MSSM fields (possibly with
  some messengers) capture all visible matter up to the GUT scale. In low scale extra dimension scenarios, 
  the degrees of freedom proliferate at  a decompactification scale $\Lambda_c\ll M_{\rm GUT}$. In  N-ification, extra degrees of freedom appear gradually via a duality cascade, that starts at $\Lambda_c$ and accelerates towards the UV. }
  \label{qmssm}
\end{figure}

In the KS gauge theory, the duality sequence unfolds evenly and the number of colors grows logarithmically with RG scale.  In general, however,  and in particular
for all known cascade extensions on the MSSM,  the duality steps tend to accumulate in shorter and 
shorter intervals, and the gauge group ranks grow at an increasingly rapid pace, as the systems flows towards the UV. 
Eventually the theory gets trapped inside a strong coupling regime and hits a so-called duality wall, located at some finite UV scale \cite{Strassler,Hanany:2003xh,Franco:2003ja,Franco:2004jz,Heckman:2007zp}. At the wall, the rank and 't Hooft coupling of the gauge theory diverge
and the gauge theory description  breaks down. In a complete framework, the physics near the
duality wall is assumed to be regulated via the decompactification transition from the 4-d gauge 
theory to a full-fledged higher dimensional string compactification.

\renewcommand{\sp}{\hspace{.5pt}}

\bigskip


\subsection*{Igniting the Cascade: Deconstructed Decompactification }

\newcommand{\Xx}{\mbox{\small $X$}}
\newcommand{\Yy}{\mbox{\small $Y$}}
\newcommand{\Zz}{\mbox{\small $Z$}}

\newcommand{\ccc}{{\mbox{\large $c$}}}

\newcommand{\qq}{{\mathsf q}}
\newcommand{\pp}{{\mathsf p}}
\newcommand{\rr}{{\bf  r}}

\newcommand{\aaa}{{\alpha }}
\newcommand{\is}{ & =  & }
\newcommand{\FF}{{\mathsf F}}

It can be shown under quite general assumptions that the embedding of the MSSM inside a cascading
gauge theory requires the introduction of one or more
extra gauge groups. The decoupling transition, at the scale
$\Lambda_c$, of this extra sector from the MSSM  can take place via three basic mechanisms \cite{Heckman:2007zp}:

\smallskip
\medskip

\noindent
${}\, $  \parbox{15.5cm}{ \addtolength{\baselineskip}{.4mm}
{\it Confinement.} The extra gauge group factor $G$ has fewer flavors than colors,
and becomes strongly coupled at some energy scale $\Lambda_c$.
Matter particles charged under $G$ confine and pair up into mesons, which manifest
themselves as MSSM matter in the~IR.}

\bigskip

\noindent
${}\, $  \parbox{15.5cm}{ \addtolength{\baselineskip}{.4mm}
{\it  Higgsing.} Bifundamental matter connecting two quiver nodes develops an expectation value at 
some scale
$\Lambda_c$, breaking a product gauge group to its diagonal. The unbroken
gauge symmetry is part of the MSSM-gauge group.
}

\smallskip
\medskip

\noindent
${}\, $  \parbox{15.5cm}{ \addtolength{\baselineskip}{.4mm}
{\it Decoupling.}  Bifundamental matter connecting two quiver nodes has a mass equal to
$\Lambda_c$. Below the scale $\Lambda_c$, the extra gauge sector is completely decoupled from the MSSM.
}

\bigskip

\noindent
A general N-ification scenario may involve a combination of these three mechanism. Moreover,
extra matter could in principle continue to show up at higher energy scales beyond $\Lambda_c$.
To keep things minimal, however, we will imagine that, once the RG cascade has been initiated, no other new gauge groups will appear. 
The UV physics then in principle follows from integrating the RG flow upwards, and by
using the rules of Seiberg duality whenever one of the gauge group factors reaches strong coupling.

\begin{figure}[t]
\begin{center}
 \epsfig{figure=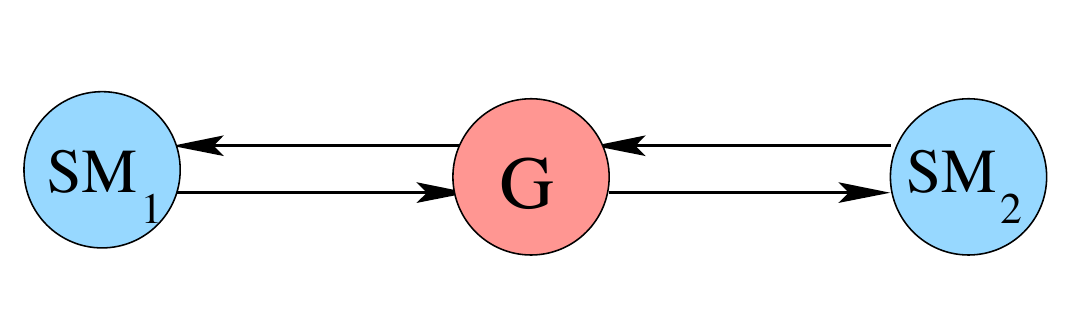,scale=0.76}
\end{center}
  \caption{The quiver diagram above the last step of the duality cascade. The gauge group $G$ has $N_f \leq N_c$ and confines
  at a scale $\Lambda_c$. The associated mesons acquire a vev that breaks the product of two Standard Model gauge groups to the diagonal.  }
  \label{qmssm1}
\end{figure}

We would like to think of the appearance of the extra matter at $\Lambda_c$ as 
are the first indication of extra dimensional physics. From this perspective, 
an attractive realization of a gauge theory one step up in the duality cascade is shown in Fig. \ref{qmssm1}. It depicts a minimal realization of the MSSM extended by
a deconstructed small extra dimension  \cite{ArkaniHamed:2001ca}.  The extra gauge group $G$ is connected via vector pairs of bifundamental matter to two copies of the Standard Model. We assume
that the number of  bifundamentals is less than the rank of $G$, so that the gauge group $G$ has $N_f\leq N_c$
and confines at some scale  $\Lambda_c$. 
The extra charged matter  pairs up to form mesons. The strong gauge dynamics generates an ADS superpotential, or a quantum modified moduli space in the case that
$N_f=N_c$. As a result, some mesons get non-zero vacuum expectation values, that, in a proper
set-up, break the 
product of two Standard Model gauge groups to the diagonal.  
Most of the mesons  acquire a mass of order $\Lambda_c$ and decouple; some may 
remain light and produce part the MSSM spectrum.

This particular realization of the first cascade step thus involves a combination of the confinement and Higgsing  
scenario. It reveals the basic characteristics of a decompactification transition of a small extra dimension \cite{ArkaniHamed:2001ca}: above the symmetry breaking mass scale set by the meson vevs,
the theory exhibits a KK doubling of the MSSM degrees of freedom. Below $\Lambda_c$ only the constant modes survives, leaving only the MSSM in the infrared.

Deconstruction produces only a rudimentary version of extra dimensional physics. In particular, it does not  include 5-d gravity. However,
once the deconstructed dimension has been created, the N-ification scenario assumes that
the rest of the extra dimensional
dynamics emerges via the RG cascade and the holographic correspondence. 
Above the deconfinement transition of $G$, the new bifundamental matter accelerates the RG running
of the two MSSM factors. Integrating upwards, the MSSM gauge groups hit strong coupling at an intermediate scale, considerably above $\Lambda_c$ but much below $M_{Pl}$.
This is where the next  Seiberg duality takes place. This process continues until at some high energy 
scale all gauge groups have large rank and large 't Hooft coupling, and the holographic AdS/CFT correspondence becomes applicable. 
The transition at $\Lambda_c$ can thus be thought
of as the first stage of a gradual decompactification process.

The energy scale above which the closed string description becomes accurate depends
on detailed parameters, such as the rank of the gauge group $G$ at the bottom
of the cascade. A priori, we may choose $G$ to have a large rank. This 
gives rise to  a low scale extra dimensional scenario.
In the following, however, we will mostly
consider the case in which  $G$ has relatively small rank, and the large $N$ regime
appears farther towards the UV -- with the wall ideally close to or somewhat below the Planck scale.

\bigskip

\subsection*{Hierarchy of Scales}

Experience shows that cascading UV extensions of the MSSM typically display an 
accelerated sequence of cascade steps that
eventually hit a duality wall at some high scale. We imagine that upon embedding
 inside a complete string compactification, the duality wall is regulated
via the transition towards the full 10-dimensional  theory.
Each specific N-ification scenario thus comes with a rather precise prediction for the hierarchy between the low 
scale $\Lambda_c$, where the new physics first shows up, and 
the high scale where the theory fully decompactifies. 
This hierarchy depends on various discrete parameters, such as the number of families and the rank of the extra gauge group~$G$.
 
Computing the hierarchy in general would seem involved or  even impossibly hard, since the cascading quiver theory typically contains one or more strongly coupled gauge sectors, so that 
anomalous dimensions are not small.  As we will show in the following, however, 
one can gain some partial control over the calculation by considering the  total beta function,
that describes the running of sum of the (inverse gauge couplings)${}^2$
\be
\label{xtot0}
x_{\rm tot} = \sum_i x_i \, , \qquad \qquad x_i = \frac{8\pi^2}{g_i^2}\, .
\ee 
This total gauge coupling
is relatively insensitive to the duality transitions, and behaves smoothly until it hits
a strong coupling singularity $x_{\rm tot} = 0$ at the location of the duality wall.
We will find that, once the theory reaches a regime where the supergravity dual description
becomes applicable, the scale dependence of $x_{\rm tot}$ becomes identified with the  profile of the dilaton along the holographic radial direction. By matching this
dilaton profile with the behavior of $x_{\rm tot}$ in the gauge theory regime, one
can make a reasonably accurate estimate of the hierarchy of scales between
the scale $\Lambda_c$ and the scale where the duality wall sets in.

Fig. 3 gives a schematic depiction of a typical profile of $x_{\rm tot}$.
Below the scale $\Lambda_c$, it follows the total coupling of the MSSM. Just above
$\Lambda_c$, the gauge theory cascade sets in. Initially, in regime I, the ranks of
the gauge groups are not yet large enough to admit an controlled string dual description.
At the intermediate range~II, all the gauge group ranks are large enough so that the
dual supergravity description becomes valid. Finally, near regime III, the duality wall appears and the
transition to the full 10-dimensional string theory takes place.

\begin{figure}[ht]
\begin{center}
 \epsfig{figure=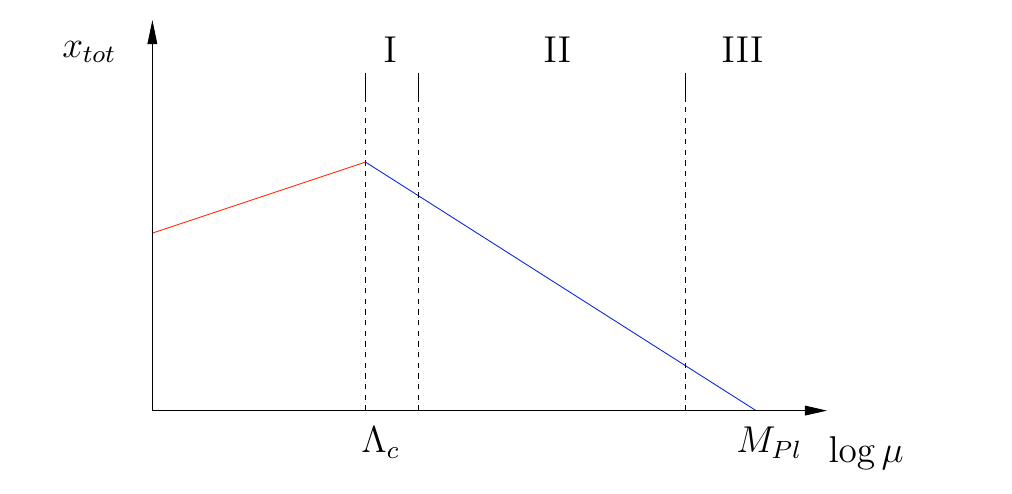,scale=1.1}
\end{center}
  \caption{The typical behavior of the total inverse gauge coupling $x_{\rm tot}$ 
  as a function of scale. The three regimes indicated are: (I) the onset of the RG cascade in the IR, (II)
  the large $N$ regime in which the dual supergravity description is valid, (III) the duality
  wall where the transition to the full 10-d string theory takes place.}
  \label{qmssm}
\end{figure}

\bigskip
\subsection*{An Explicit Example}

The N-ification program can in principle be formulated and developed from a purely
field theoretic perspective. Indeed, one may view it as an approach to 
string phenomenology that, rather than looking for ways to derive particle physics from
string theory, looks to develop scenarios for which the UV string theory can be derived from the field theory. To find geometrically 
attractive realizations, however, it is natural to look for guidance from string theory.
So we adopt the hypothesis that the Standard Model, and its UV extension in the form
of a cascading
gauge theory, both arise as the low energy effective field theory of open strings attached 
to a configuration of D-branes, placed within some suitable geometry.

As in \cite{Klebanov:2000hb}, we are interested in finding periodic examples,
for which the quiver diagram returns to itself after a few Seiberg duality steps, but with 
a new assignment of  gauge groups ranks. For such periodic cascades, the 4-d quiver 
gauge theory reaches a smooth large $N$ limit, for which it may be easier to identify a dual 
string description.

An interesting cascading extension of an MSSM-like theory
was introduced and studied in \cite{
Cascales:2005rj}. The quiver theory
has the structure shown in Fig. 2, where upon evolving the cascade towards the UV, the two light-blue nodes turn 
into large $N$ versions of the Standard Model gauge group
$U(N+3) \times U(N+2) \times U(N+1)\, .$ The model was shown to admit an explicit string theory realization in terms of D3 and D7-branes, placed on a $\Zbb_3$
orbifold of a particular Calabi-Yau singularity, known as the cone over the suspended pinch point. 
The theory has the hierarchical structure shown in Fig. 3. In the large $N$ region II, it has  
a known supergravity dual that develops a duality wall near the Planck scale.\footnote{Near the IR end of region II, energies are comparable to $\Lambda_c$. Because of this, the
effect of the complex deformation dual to the strong dynamics of $G$ is important. Despite being conceptually equivalent
to KS, finding an explicit gravity dual for this region (based on the deformed cone) is formidable task due to the
small isometry group. In section 6, we present the gravity dual based on the singular cone, which is valid at energies
well above $\Lambda_c$. This solution is analogous to the Klebanov-Tseytlin one for the conifold \cite{Klebanov:2000nc}.} In the infrared, it reduces to a semi-realistic
field theory, that shares most essential features (gauge group,
chiral matter content, and couplings) of the MSSM. We will center our presentation 
on this particular example, as it provides a useful and attractive illustration of our general ideas.\footnote{Our choice of flavor D7-branes differs slightly from the one in \cite{Cascales:2005rj}, but their analysis applies with trivial modifications.} We would like to reiterate, however, that our general perspective can also be applied to pure
field theoretic models, that may reproduce low energy phenomenology in greater detail.

\bigskip

\subsection*{Outline of the paper}

This paper is organized as follows. In sections 3 to 4, we review the particular 
embedding of the MSSM quiver theory inside of duality cascade introduced in \cite{
Cascales:2005rj}. 
Our exposition will be somewhat schematic, since most technical geometric
aspects of the model are carefully worked out and explained in the original paper \cite{
Cascales:2005rj}.
We will attempt to motivate the construction and highlight various physical aspects.
The quiver theory associated with an SPP singularity is introduced in section 2, as an
example of a deconstructed small extra dimension. The
MSSM-like quiver theory an its D-brane realization is summarized in section 3;
some more detailed discussions are referred to appendix A. Sections 4 and 5 study the RG cascade. We discuss the calculation of the running of gauge couplings in strongly coupled cascading theories and present the computation of the hierarchy of scales
between the onset of the cascade and the location of the duality wall.  We will find that the UV scale is roughly 12 orders of magnitude
above $\Lambda_c$.

In section 6 we explain some general properties of supergravity duals of D-brane gauge theories
with flavor D7-branes. In particular, we explain how the presence of the
D7-branes naturally leads to an accelerated RG cascade, with a
non-trivial dilaton profile and a duality wall.  We compute the total beta-function,
that governs the radial dependence of the dilaton, and show that it is proportional 
to the number of  flavor branes.  The precise relation naturally follows from the fact that  
the dilaton is paired with the axion, which in turn is sourced by the D7-branes. We show that the supergravity
result agrees with the gauge theory calculation.

\bigskip


\section{A Deconstructed Small Extra Dimension}


\label{section_deconstructed_dimension}

In this section, we present the geometric realization of a three node quiver, that takes the form of 
a deconstructed extra dimension of the type shown in Fig. 2. As a warm-up,
we first consider the toy example in which the Standard Model node is replaced 
by a simple $U(N)$ gauge factor. 
The set up then involves $N$ D-branes placed at a simple Calabi-Yau singularity known
as the suspended pinch point (SPP) singularity.

\subsection{D-branes at the Suspended Pinch Point}

When placed in the proximity of a Calabi-Yau singularity, D-branes typically rearrange themselves
into so-called fractional branes. Geometrically, these fractional branes can be thought of as bound states 
of several D-branes, each wrapped around small compact cycles supported within the singularity. The number of different types of fractional branes equals the number of independent compact cycles. 
The gauge theory on a collection of fractional branes takes the form of a quiver gauge
theory, where each stack of $N$ fractional branes represents a separate $U(N)$ quiver node.
The open strings at the intersections between the branes give rise to 
massless bifundamental matter, represented by the oriented lines that connect the corresponding quiver nodes.

The suspended pinch point (SPP) singularity may be obtained via a partial resolution of a $\Zbb_2\times \Zbb_2$ singularity  \cite{Morrison:1998cs}. It
is described by the following
 equation in $\Cbb^4$
\be
\label{spp}
\qquad cd=a^2b, \qquad \qquad \quad \mbox{\small $(a,b,c,d) \in \C^4\, .$}
\ee
Geometrically, the singularity looks like a complex cone over a 4-d base manifold $\Sigma_4$.
For our present purpose, it is sufficient to know that the singularity supports three types of fractional
branes, and thus gives rise to a quiver theory with three nodes.
The quiver  for a specific collection of fractional branes on the SPP singularity
is shown in Fig. 4. The two light-blue nodes indicate $U(N)$ gauge factors. The superpotential
of this SPP quiver gauge theory takes the form
\be
\label{super}
W\! =\Tr \bigl(\Phi( \mbox{\small $X\tilde X-\tilde Y Y)
- Z \tilde Z \, \tilde X X + \tilde Z  Z  Y \tilde Y$}  \bigr)
\ee
The two nodes without adjoints are fractional branes that wrap rigid cycles; the node with the adjoint
corresponds to a brane that is free to move in one direction.

\begin{figure}[t]
\begin{center}
 \epsfig{figure=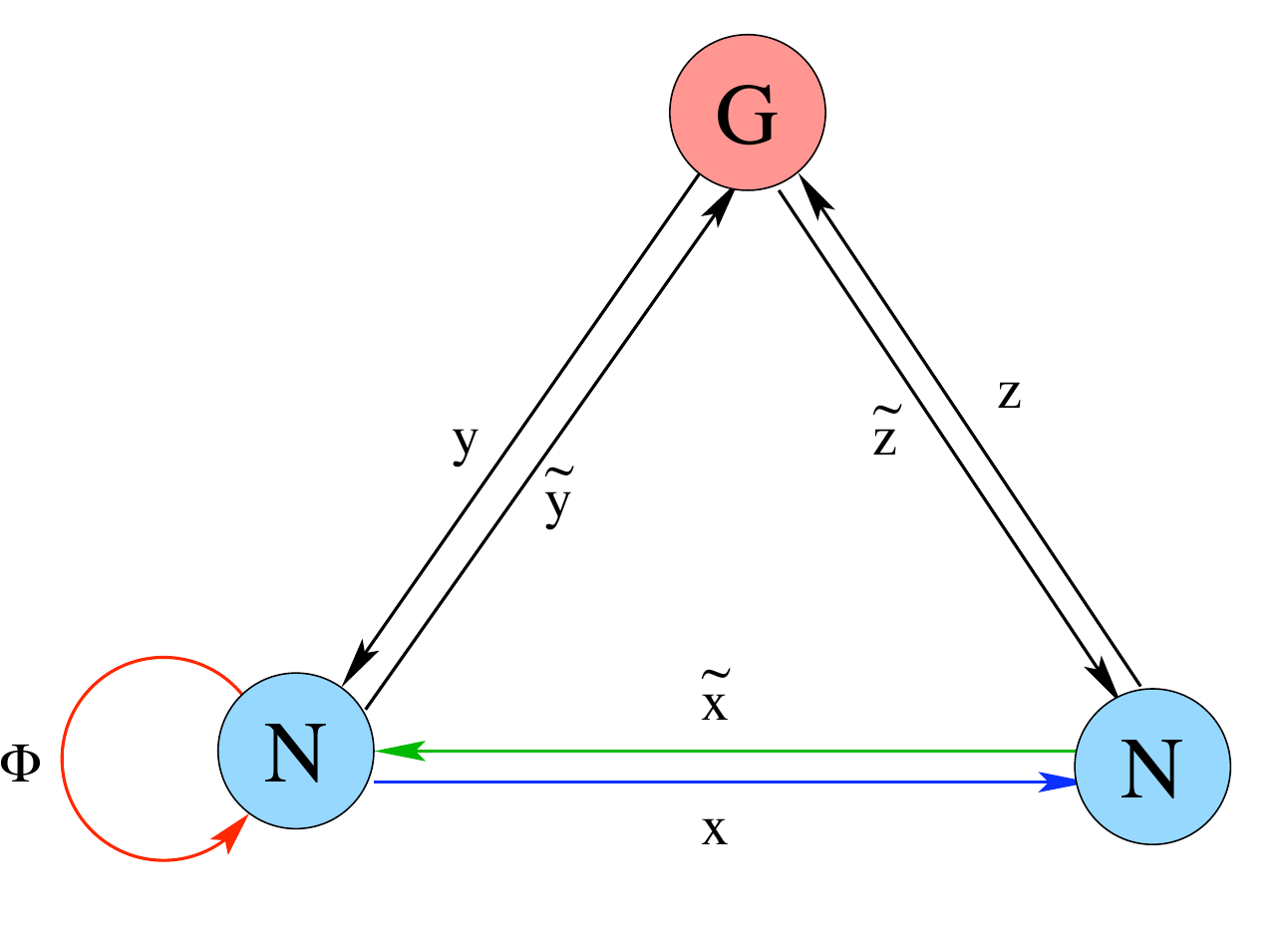,scale=0.6}
\end{center}
  \caption{The three node quiver of N D3-branes and $M$ fractional branes on an SPP singularity. 
The extra node  with gauge group $G = U(N\! +\! M)$ with $M >\! N$ confines in the~IR.}
\end{figure}

The rank of a quiver node can be adjusted by changing
the multiplicity of the corresponding fractional brane. 
In case the third group $G$ has the same rank as the other two nodes, the quiver represents the
world volume gauge theory of $N$ D3-branes on the SPP singularity and flows to a conformal fixed point in
the IR. In the large $N$ limit, this CFT has a known supergravity dual.\footnote{We will review some properties of
this supergravity dual in section 5.  }
To confirm the geometric origin of this quiver theory, one may verify that its moduli space, 
given by the space of expectation values of the chiral
matter fields modulo gauge transformations and F- and D-term equations,
is equal to the configuration space of $N$ D-branes on the SPP geometry.\footnote{
For the case that $N=1$, one can readily verify that
the gauge invariant combinations $a=$ {\small $\tilde XX$} = {\small $\tilde YY$}, 
$b=$ {\small $Z\tilde Z$}, 
 $c= $ {\small $ \tilde X\tilde Y\tilde Z$} and $d= $ {\small $Y X Z$} 
satisfy the SPP equation (\ref{spp}).}

To realize the deconstructed extra dimension of the type shown in Fig. 4, we choose the
 multiplicity $M$ of the third fractional brane 
such that the third node $G$ has rank $N+M > 2N$. In this case,  it will flow to strong coupling
towards the IR and confine at some scale $\Lambda_c$. The chiral matter fields charged
under $G$ combine into mesons $M_{\rm yy} =$ {\small $\tilde Y Y$}, $M_{\rm yz} =$ {\small $\tilde Y \tilde Z$}, etc. The strong coupling dynamics gives rise to an ADS contribution to the superpotential.
The superpotential  then takes the form
\bea
W \is  \Phi (\mbox{\small $X \tilde X$}-M_{\rm yy})- M_{\rm zz}\mbox{\small $\tilde X X$}+ M_{\rm yz}M_{\rm zy}
\; + \; (M-N)\left(\frac{\Lambda^{3M+N}}{\det{\cal M}}\right)^{\frac{1}{M-N}}
\eea
with
\bea
{\cal M} = \left[\begin{array}{cc} M_{\rm yz} & M_{\rm yy} \\[1mm]  M_{\rm zz} & M_{\rm zy} \end{array} \right] = \left[\, \begin{array}{cc} \tilde\Yy \tilde\Zz \, & \,   \tilde\Yy \Yy \\[2mm]  \Zz \tilde\Zz\, & \, \Zz \Yy \, \end{array}\right] \, .
\eea
The analysis simplifies slightly for the case $N=1$. The equation of motion of $M_{\rm zy}$ sets $\det{\cal M}=\Lambda^{3M+1 \over M}$. The mesons  $M_{\rm yy}$, $M_{\rm yz}$, $M_{\rm zy}$, and the linear
combination $\Phi -  M_{\rm zz}$ acquire a mass.
On the branch where only $M_{\rm yz}$ 
and $M_{\rm zy}$ have non-zero vevs, the two light-blue nodes collapse to
the diagonal $U(N)$ gauge theory, with three remaining light adjoint matter fields {\small $X$, $\tilde X$} and
$\Phi + M_{\rm zz}$.\footnote{The massless combination of $\Phi$ and $M_{\rm zz}$ carries some power of $\Lambda$ for generic $N$.} By interpreting the two $U(N)$-nodes as two points inside a small deconstructed
extra dimension, the confinement transition may be viewed as the KK-reduction to the 4-d theory.

The quiver with $G=U(N+ M)$ describes the world-volume theory of $N$ D3~branes and $M$
fractional branes, given by D5-branes that wrap a compact 2-cycle within the
base of the SPP singularity. This set-up is analogous to the Klebanov-Strassler system,
and similarly leads to a cascading gauge theory with holographic dual given by a deformation
of the orginal SPP singularity supported by fluxes \cite{Franco:2005fd}. The passage from the 
D-brane theory to the smooth supergravity dual is a brane/flux geometric transition,
that replaces the $M$ fractional branes by $M$ units of RR 3-form flux over finite size 3-cycles. 
The deformed geometry
is described by the equation
\be
\label{deformed}
cd-a^2b = \epsilon a.
\ee
where $\epsilon$ is the deformation parameter. This geometry can be recovered 
by studying the mesonic moduli space after the confinement 
transition at $\Lambda_c$. 
The mesons represent the location of the remaining $N$ D3-branes, moving inside the
deformed SPP. 
The supergravity description of the
deformed geometry, however, becomes accurate only if the rank of the confining node $G$ is sufficiently large. The correspondence between gauge theory dynamics and geometry was investigated in great detail in \cite{Franco:2005fd} and, in a context closer to this paper, in \cite{Cascales:2005rj}. Many examples, including our main case of study, were presented in these references. Thus, we refer the reader to them for details.


In the above toy example,
the deformed geometry is smooth. 
If the geometry after the deformation remains singular, a chiral gauge theory such as the MSSM can arise on the worldvolume of the D3-branes. 
In the next two sections, following \cite{Cascales:2005rj}, we will review how this toy example can be generalized into a semi-realistic 
UV extension of the MSSM. In this generalization, the
two light-blue nodes each host  one copy of the MSSM gauge group. 
Moreover, we will see that the three remaining light fields  {\small $X$, $\tilde X$} and
$\Phi + M_{\rm zz}$ will carry the correct quantum numbers to be identified with
three MSSM matter generations in the IR.

\bigskip

\section{An MSSM-like quiver}

Because the MSSM takes active part in the duality cascade, we need to embed it inside
a quiver theory  with somewhat special characteristics that
allow the gauge group ranks to increase with every cascade step, without changing the quiver data.
These special properties are naturally implied from the geometric rules of D-brane engineering.

The specific MSSM-like quiver theory that we will use as our main example is depicted in Fig.~5. The three
light-blue colored nodes represent the three Standard Model gauge groups, extended to 
\be
\label{smgroup}
U(3) \times U(2) \times U(1).
\ee
The model has the same chiral matter content, indicated by the oriented lines,  
as the MSSM, three generations of quarks 
and leptons (each generation has its own color),
except that it has a non-minimal Higgs sector, one pair of Higgs doublets per generation.  The right-handed neutrinos are also missing.
The theory is free of non-abelian anomalies but has mixed $U(1)$/non-abelian anomalies. Only the hypercharge
combination
\beq
Y=-Q_1-{Q_2 \over 2}-{Q_3 \over 3} \, 
\eeq
of abelian gauge symmetries is anomaly free and remains light. The remaining two anomalous $U(1)$ gauge bosons are assumed to acquire a large mass, of order the string scale, via the familiar Green-Schwarz anomaly cancellation mechanism \cite{Green:1984sg}.

\begin{figure}[hbtp]
\begin{center}
 \epsfig{figure=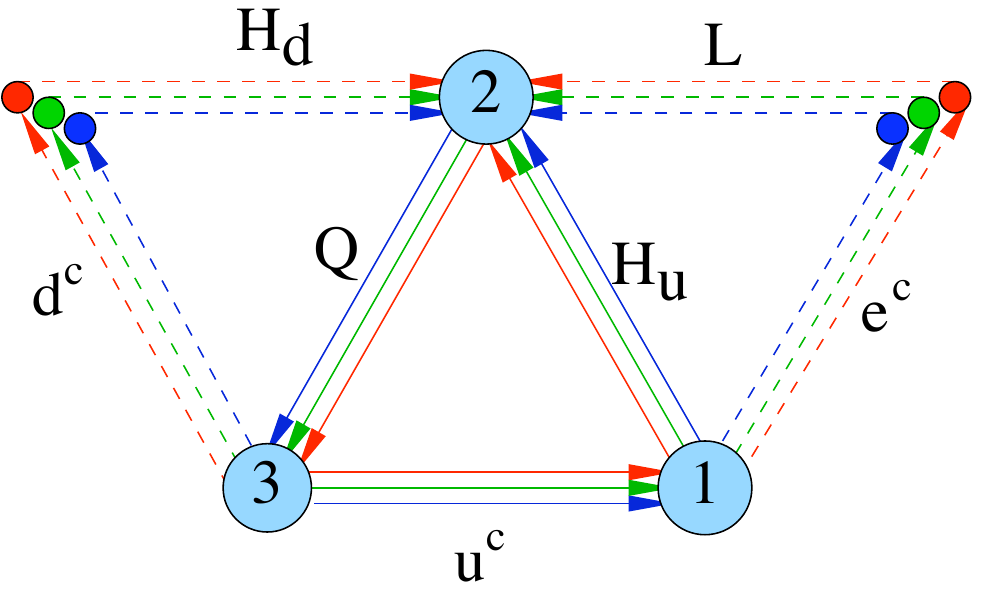,scale=.83}
\end{center}
  \caption{The MSSM quiver theory obtained from  $\Cbb^3/\Zbb_3$ with
  six fractional D-branes (light-blue nodes)
  and six flavor D7-branes. It describes an anomaly free
  gauge theory, with the matter content of the MSSM plus two extra pairs of Higgs doublets.}
  \label{qmssm2}
\end{figure}

The small colored vertices on the upper left and right corner represent flavor nodes.
They do not support a local gauge
symmetry group. We can distinguish
two different types. The three nodes
on the left support chiral matter charged under $U(3) \times U(2)$; the corresponding
lines are oriented such that the superpotential contains cubic Yukawa couplings ${\Tr}(\tilde d \, H_d \, Q)$ of the down-type quarks. The  lines connected to the three flavor nodes on the right, on other other hand, are oriented in the opposite direction, forbidding the presence of cubic terms
in the superpotential. Rather than with cubic Yukawa couplings, 
the superpotential starts out with quartic terms of the form ${\Tr}(\tilde e \, L\, Q\, \tilde u)$. 

The quiver  theory of Fig. 5 can be engineered by
placing D-branes on a $\Cbb^3/\Zbb_3$ singularity. The brane configuration involves
6 so-called fractional branes,
that wrap small compact cycles within the orbifold geometry, and
give rise to the local  gauge symmetry (\ref{smgroup}). The  flavor nodes arise
in the geometric construction from
D7-branes that wrap non-compact cycles. The main elements of the construction
are summarized in Appendix A; a more detailed exposition is found in the original literature \cite{Aldazabal:2000sa}.\footnote{See also \cite{Berenstein:2001nk,Alday:2002uc,Verlinde:2005jr} for other realizations of MSSM-like theories with D-branes on singularities.}

The low energy phenomenology of the  MSMM-quiver model depends on the precise values
of the various couplings, on the mediation mechanism for supersymmetry breaking, etc. 
We will not concern ourselves with these questions here, except to note that,
as a world-volume theory of a local brane construction, all gauge invariant couplings (gauge
and Yukawa couplings, soft parameters, etc) can in 
principle be freely adjusted by tuning local geometric data. Variations of the superpotential correspond to complex structure deformations of the $\Cbb^3/\Zbb_3$
geometry,  gauge couplings can be modified by turning on 2-form fields, and soft
parameters arise by turning on F-term components of  closed string fields. 

The quiver MSSM theory of Fig. 5 has the important feature, common to most D-brane world-volume 
 theories, that it allows for a direct generalization to a consistent anomaly-free large $N$ field theory,
 without changing the quiver data. If we replace the gauge group 
 by \be
U(N+3) \times U(N+2) \times U(N+1),\ee
while keeping the number of flavor nodes fixed, the theory remains free of non-abelian anomalies.
The N-ification scenario of \cite{Cascales:2005rj}, that we will present below, makes essential use of this property.

\bigskip

\section{The Duality Cascade}

Still following \cite{Cascales:2005rj}, we now put the various ingredients together to
obtain a geometric realization of a cascading extension of the MSSM-like gauge theory.

At the field theory level, the plan is to construct a gauge theory with the schematic 
quiver structure depicted below in Fig.~6. It shows the SPP quiver gauge theory,
with the extra flavor nodes added `by hand'. The two light-blue nodes
  represent two gauge groups that, at the bottom of the cascade, reduce to two 
  copies of the Standard Model gauge group. 
  The small colored nodes are the flavor nodes and the  
  colored lines are chiral fields with the quantum numbers of the
3 generations of quarks and leptons relative to the  left SM node, 
  except that the two generations of  (up)quarks are in bifundamental of the left and right SM node.
 In approaching the IR, the node 
  $G$ has $N_f<N_c$ and confines at the scale~$\Lambda_c$.   The corresponding mesons
  acquire vevs, similar as discussed for the toy model
  in Sect. 2. 
  The challenge is to find the correct structure for the gauge group $G$ and bifundamental
  matter such that below the confinement transition, the effective field theory that remains in the infrared has the gauge symmetry and field content of the MSSM.   

This field theoretic puzzle has a beautiful geometric solution, obtained by combining the lessons
of the two previous sections  \cite{Cascales:2005rj}.
 The deformed SPP geometry, as specified by eqn. (\ref{deformed}),
is invariant under the $\Zbb_3$ action 
\be
\label{zthree}
a \to \omega^2 a\sp , \; b \to \omega\sp b\sp ,\; c \to \omega\sp c\sp ,\; d \to \omega \sp d
\ee
with $\omega^3 = 1$. It is therefore possible to define the $\Zbb_3$-orbifold of the
deformed SPP~singularity. Thanks to the deformation, the local region 
near the fixed point locus of the $\Zbb_3$ action looks like the local region near
the fixed point of the flat orbifold $\Cbb^3/
\Zbb_3$. We can thus place the same combination of fractional branes and D7-flavor branes
on the deformed SPP orbifold,  that produces the MSSM quiver of Fig.~5 when placed on the 
flat orbifold. The holographic correspondence then virtually guarantees that the resulting 
gauge theory will have the quiver structure as indicated in Fig.~6, and will exhibit an RG
cascade that flows  to the MSSM-quiver theory in the infrared. 

\begin{figure}[t]
\begin{center}
 \epsfig{figure=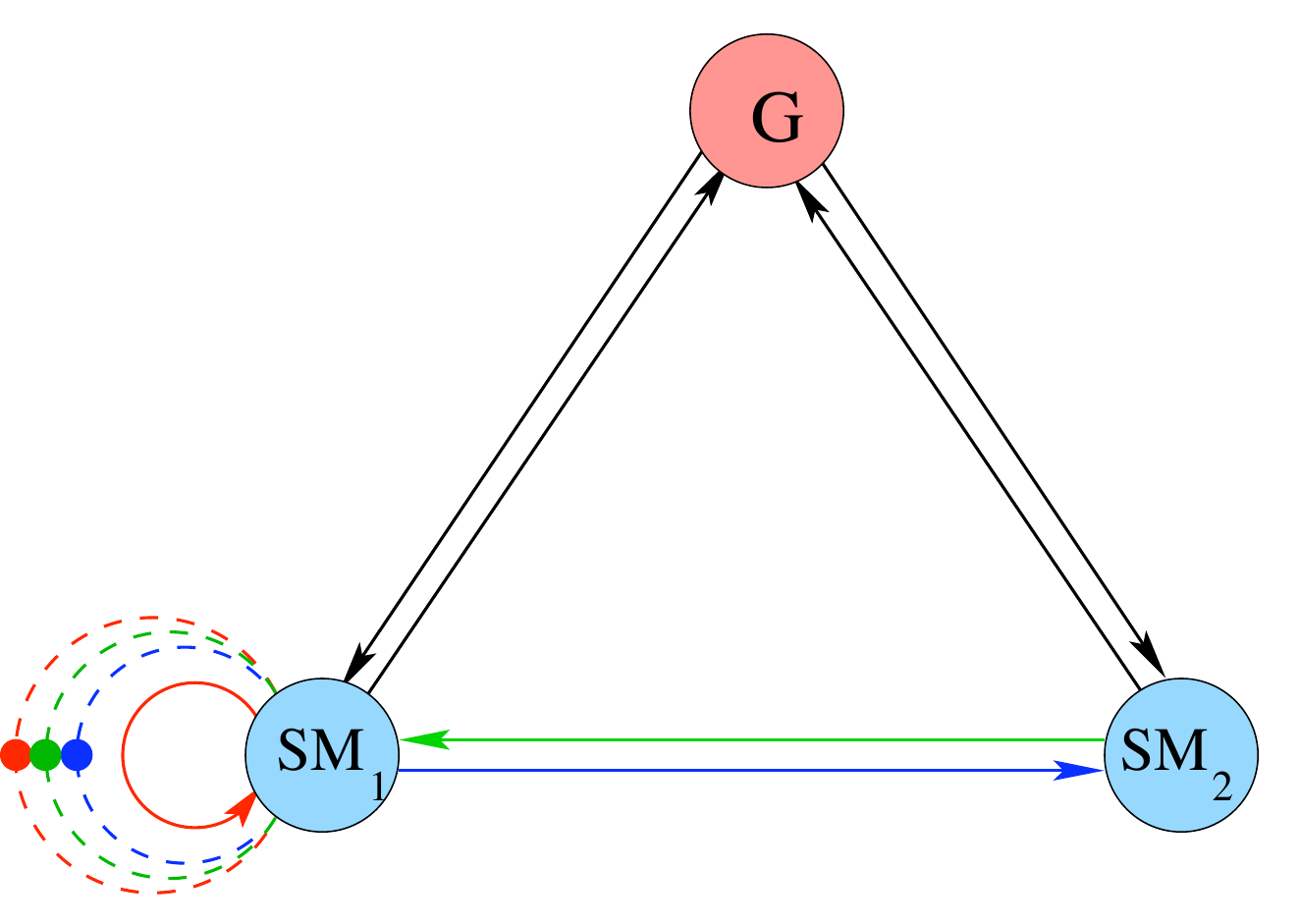,scale=0.6}
\end{center}
  \caption{Schematic quiver of the cascading field theory. The light-blue nodes
  represent~two copies of the Standard Model gauge group. The third node $G$ confines
  in the IR, leading~to~vevs for the mesons that break the gauge symmetry to that of the Standard Model.
  The colored lines are chiral fields that in the IR reduce to 3 generations of quarks and~leptons.}
  \label{qmssm}
\end{figure}

\medskip

\subsection{Bottom of the Cascade}

The SPP$/\Zbb_3$ geometry is attractive,
 because the corresponding gauge theory can be easily derived by orbifolding that for the SPP. By now it is known how to derive the field theory on D-branes on arbitrary toric singularities using dimer model techniques \cite{Hanany:2005ve}-\cite{Franco:2007ii}. 
Schematically, we know that the orbifold projection will split each of the three nodes of the 
SPP gauge theory into a product of three gauge groups, corresponding to the three irreducible
representations of $\Zbb_3$.  In \cite{Cascales:2005rj}, the Chan-Paton factors are chosen as follows
\bea
\label{chanp}
&\gamma_{{}_{\rm SM1}} & = \ {\rm diag} \, (\, \onebb_1\, ,\; \omega\sp \onebb_2\, ,\; \omega^2\sp \onebb_3\, )\nonumber\\[2mm]
& \gamma_{{}_{\rm SM2}} & =  \ {\rm diag} \, (\, \onebb_2\, ,\; \omega\sp \onebb_3\, ,\; \omega^2\sp \onebb_1\, )\\[2mm]
& \gamma_{G} & = \, {\rm diag}\sp (\sp \onebb_{M+3},\, \omega\sp \onebb_{M+1}, \, \omega^2\sp \onebb_{M+2}), \nonumber
\eea
which, as promised, breaks the gauge symmetry to two copies of the Standard Model gauge group,
times the extra gauge factor 
\be
G = U(M+1) \times U(M+2) \times U(M+3).
\ee
The $\Zbb_3$  transformation rule of the chiral matter fields  follows from (\ref{zthree}) and
the identifications given in the second footnote in Sect.~2. Using the  notation as in Fig.~4
\bea
\Xx \to \omega^2 \Xx \  \ \ ;  \ \ \   \tilde\Xx \to \tilde \Xx \ \  \ ; &  & \  \Yy \to \omega^2 \Yy \ \ ;  \ \  \tilde\Yy \to  \tilde \Yy \ \ ;  \  \  \nonumber \\[-2mm]
\\[-2mm]
\Zz \to  \Zz \ \ ;  \ \ \  \tilde\Zz \to \omega \tilde \Zz & &  ;  \  \  \  \Phi \to \omega\sp \Phi \nonumber
\eea
Together with the chosen assignment of Chan-Paton factors (\ref{chanp}), these transformation rules dictate how the chiral matter lines connect  between the quiver nodes: a field connects
only those pairs of nodes for which the relative Chan-Paton phase cancels its $\Zbb_3$ charge. In a
similar way, one may assign the proper Chan-Paton phases to the D7-branes, so that the corresponding
flavored chiral matter fields connect to the Standard Model nodes in accordance with the
MSSM-quiver of Fig.~5. 

\begin{figure}[t]
\begin{center}
 \epsfig{figure=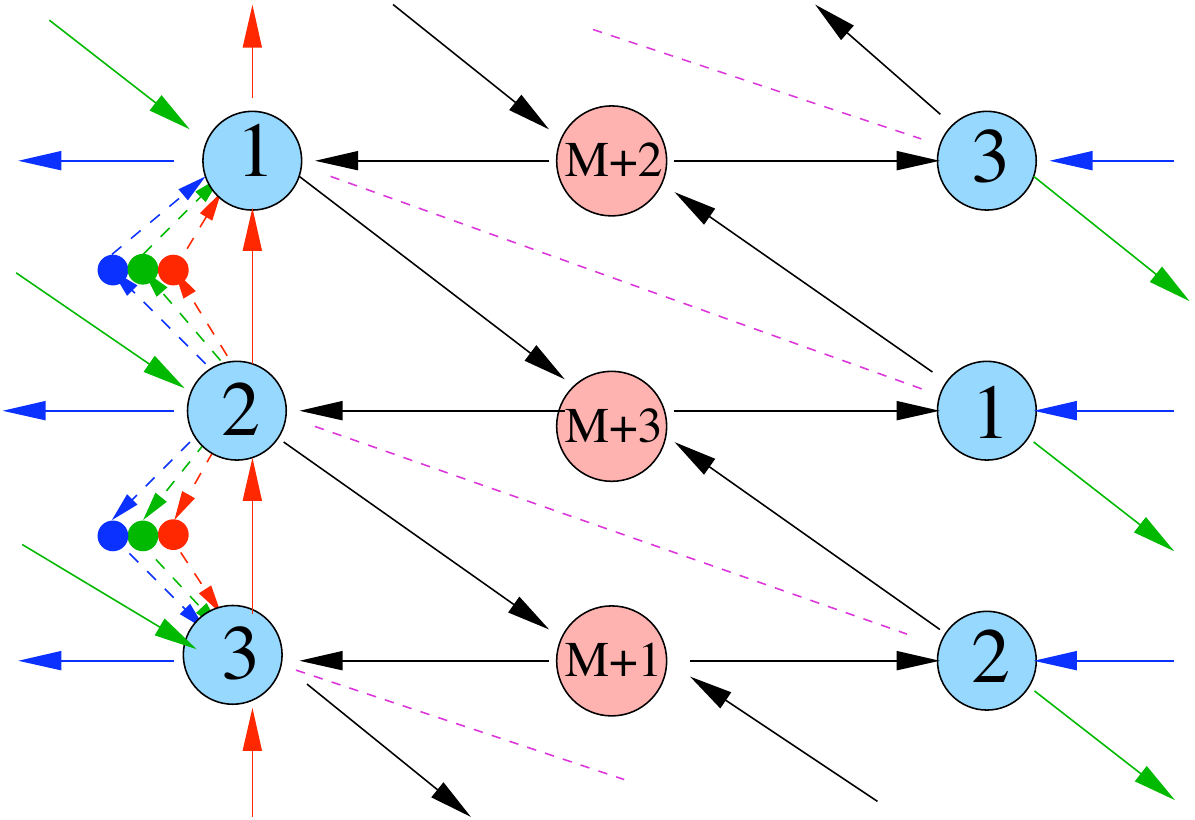,scale=.7}
\end{center}
  \caption{The quiver diagram of the SPP/$\Zbb_3$ orbifold gauge theory with flavor nodes.
 The mesons that acquire a non-zero vevs upon confinement of $G$ are lightly indicated in magenta.}
  \label{qmssm}
\end{figure}
\newcommand{\Vfr}{\mathfrak{v}}
\newcommand{\halff}{\mbox{\large $\frac{1}{2}$}}

The resulting quiver diagram is shown in Fig.~7 (periodic identification of the horizontal and vertical directions is understood). Note that all nodes have an equal number of incoming
and outgoing lines, ensuring that all non-abelian anomalies cancel.
We have also indicated, with light dashed magenta lines, the
mesons that will acquire an expectation value after confinement of the three $G$ nodes in 
the middle column. 

The spectrum of light particles and the symmetry breaking pattern after the confinement transition 
 is found by following the same steps as outlined for the toy model in Sect.~2.
The superpotential of the orbifold gauge theory 
is inherited from the cover theory, and thus looks identical to eqn. (\ref{super}),
except that all chiral fields must be generalized to carry two indices that label the two nodes that they connect~to.
The analysis of the orbifold spectrum is described in detail in \cite{Cascales:2005rj}, and we will not repeat it here. 

The result one finds is that, among the
mesons formed out of the chiral matter charged under $G$ (the black arrows in Fig.~7)
only the mesons $M^{ii}_{\rm yz}$ and $M^{ii}_{\rm zy}$ that connect left and right SM nodes 
with the same rank acquire a vev.
This leaves the diagonal SM gauge group unbroken.
As in Sect.~2 (and using the same notation) one finds that the mesons  $M^{ij}_{\rm yy}$, 
$M^{ij}_{\rm yz}$, $M^{ij}_{\rm zy}$, and the linear combination $\Phi^{ij} -  M^{ij}_{\rm zz}$ acquire a mass.
The matter fields that remain light are {\small $X^{ij}$, $\tilde X^{ij}$} and $\Phi^{ij} + M^{ij}_{\rm zz}$, and all the flavored matter connected to the flavor nodes. 

In other words, all colored lines in Fig.~7
directly reduce to light particles, except that the solid red arrows (the $\Phi$ fields)
mix with the meson fields $M^{ij}_{\rm zz}$, that are charged under the right copy of the SM gauge
group. Physically, this means that all SM particles in this model are elementary
(at least relative to this last confinement transition at the bottom of the RG cascade) except that
one generation of the left-handed quark doublet and right-handed up quark is partially
composite. 
From a phenomenological perspective, it is natural to identify the two elementary quark families
with the first two generations, and the partially composite quarks with the third generation.\footnote{
Experimental bounds, from tests of $Z\to b \bar b$, only allow for a small amount of TeV-scale 
compositeness  of $b_L$. As we will discuss later, we have other reasons to assume that
the confinement scale $\Lambda_c$ will be in the multi-TeV range.}


\medskip



\subsection{The SPP/$\Zbb_3$ Cascade}


The duality cascade for SPP$/\IZ_3$ was analyzed in detail in \cite{Cascales:2005rj}. Starting from Fig. \ref{qmssm2} at some UV point, it terminates in Fig. \ref{qmssm} as $N$ is gradually reduced to zero. Similarly, $M$ also decreases towards
the IR. The cascade follows directly from the one in the parent SPP theory, which at each step dualizes the highest
rank node out of the two without an adjoint \cite{Franco:2005fd}. After each dualization, the quiver becomes again the SPP one, with the adjoint 
field shifted to another node. Seiberg duality on a single SPP node translates to a dualization of all three nodes in a
column of the SPP$/\IZ_3$ quiver. The cascade then corresponds to sequentially dualizing the three columns.

After each dualization, the D7-branes `move' over the quiver, i.e (anti)fundamental fields connect to new nodes \cite{Cascales:2005rj}. 
When a node that contains (anti)fundamental fields is dualized, some of the resulting Seiberg mesons are singlets under all the gauge groups. These fields transform in bifundamental representations of the global symmetry nodes in the extended quiver and can be represented by arrows connecting them. These mesons, together with their evolution with scale, holographically encode degrees of freedom in D7-D7' sectors. They are generically necessary in flavored cascades in order to match anomalies in global symmetries. In this particular example, they become massive after a few steps and can be integrated out. In other cases, such as in the model studied in \cite{Benini:2007kg}, their number can build up towards the UV.

\begin{figure}[hbtp]
\begin{center}
 \epsfig{figure=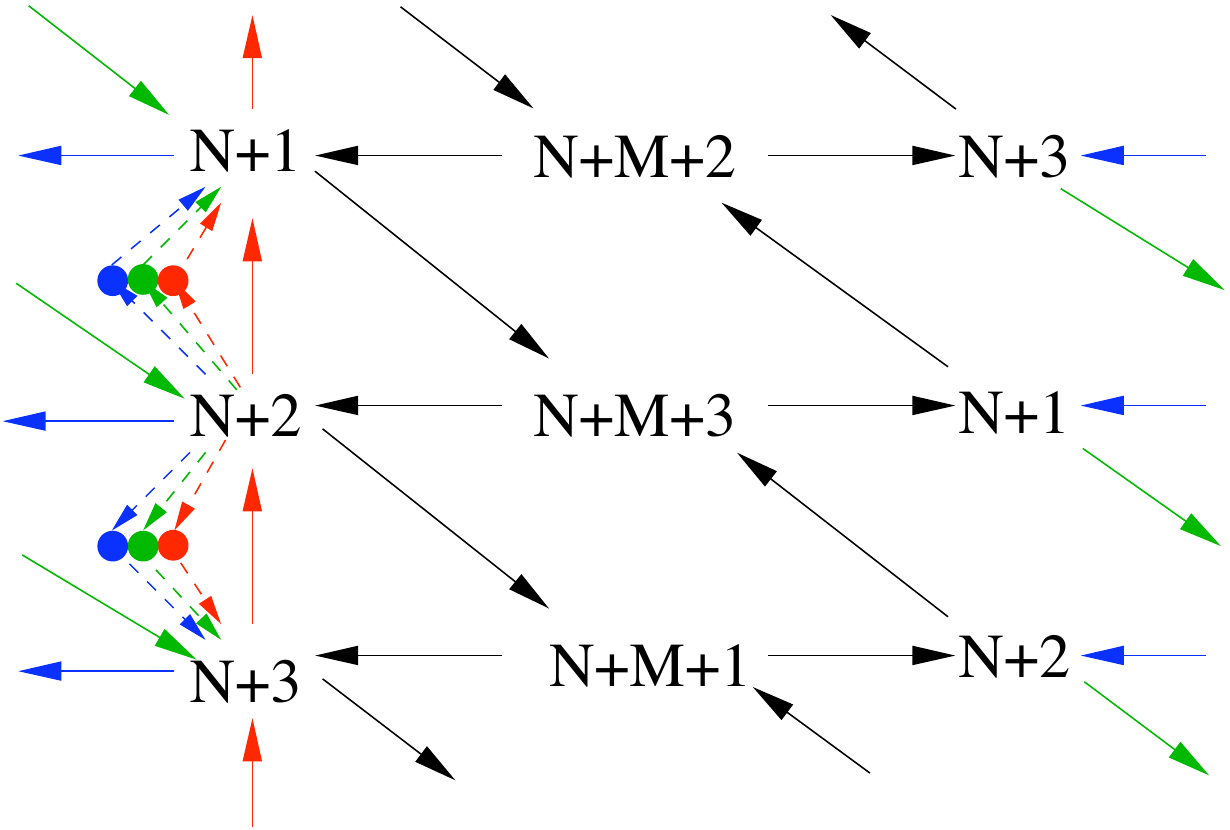,scale=.65}
\end{center}
  \caption{The quiver diagram of the SPP/$\Zbb_3$ orbifold gauge theory with flavor nodes, at some arbitrary point in the cascade.}
  \label{qmssm2}
\end{figure}


\section{RG running}

Let us review the basic ideas necessary for computing duality cascades from a field theory perspective. Cascading theories are generically at strong coupling, with fields having ${\cal O}(1)$ anomalous dimensions. It is crucial to take into account these large anomalous dimensions (equivalently superconformal R-charges) to appropriately determine the running of couplings. The starting point is a superconformal quiver gauge theory with $\prod_{i=1}^k SU(n_i)$, with $n^i=r^i \,N$.\footnote{Most of the examples considered in the literature correspond to $r_i=1$ for all $i$.} To compute the anomalous dimensions in the superconformal theory, we first require the vanishing of the beta functions for gauge couplings
\beq
\beta_i={dx_i \over d \ln \mu}=3 \,n^i +{1\over 2} \sum_{j=1}^k (f_{ij}(\gamma_{ij}-1)+f_{ji}(\gamma_{ji}-1))n^j
\label{beta_g}
\eeq
with $x_i$ given in \eref{xtot0}. We present equations in a form that is most suitable for its application to quiver gauge theories. $f_{ij}$ is the symmetric adjacency matrix of the quiver and gives the number of bifundamental multiplets between nodes $i$ and $j$. The superpotential is of the form $W=\sum_\mu h_\mu W_\mu$, with $W_\mu$ gauge invariant monomials.
The couplings $h_\mu$ have scaling dimension
\beq
\beta(h_\mu) = -d(h_\mu)+{1\over 2} \sum_{X_{ij}\in W_\mu} \gamma_{ij}
\label{beta_h}
\eeq
where $d(h_\mu) = {\rm deg}(W_\mu)-1$ is the naive mass dimension of $h_{\mu}$.

The discussion can equivalently be casted in terms of superconformal R-charges, which can be traded by anomalous dimensions by means of the usual relation $R_{ij}=2/3 + \gamma_{ij}/3$. Typically, the equations above do not fully determine the anomalous dimensions. This is a manifestation of the fact that conformal invariance is not sufficient to rule out the possible mixing of the superconformal R-charge with additional $U(1)$ global symmetries. The remaining freedom
is fixed by a-maximization \cite{Intriligator:2003jj}. Within the space of solutions to \eref{beta_g} and \eref{beta_h}, the R-charges must maximize the trial central charge

\beq
a={3\over 32}\left[2 \sum_i n_i^2 + \sum_{i<j} f_{ij}n_i n_j [3(R_{ij}-1)^3-(R_{ij}-1)]\right]\ .
\eeq

Conformal invariance is then broken by introducing small numbers (compared to $N$) $M$ and $K$ of D5 and D7-branes.\footnote{Our discussion generalizes without changes to cases with more than one type of D5 and D7-branes.} From a field theory perspective, this corresponds to small modifications of the gauge group ranks and a small number of (anti)fundamental fields.

In the presence of D5 and D7-branes, anomalous dimensions have an expansion that schematically looks like

\beq
\gamma=\sum_{m,n=0}^{\infty}\gamma^{(m,n)} (M/N)^m (K/N)^n \, .
\eeq
We will be later interested in computing the beta fuction of the total coupling \eref{xtot0}, which we denote $\beta_{tot}$. In order to determine it to leading order, it is sufficient to consider the conformal values $\gamma_{i,j}^{(0,0)}$ (equivalently the associated R-charges). We justify this statement below.

When computing the beta functions for individual gauge couplings, we must add to \eref{beta_g} the contributions of the additional fundamental flavors. Combining all the beta functions, we obtain

\begin{eqnarray}
\beta_{tot}&=&\beta_{tot}^0+{3 \over 2}\sum_{D7_i}(R_{q_i}+R_{\tilde{q}_i
}-2) \nonumber \\
&=&\beta_{tot}^0-{3 \over 2}\sum_{D7_i}R_{X_i}
\label{beta_extended}
\end{eqnarray}
$\beta_{tot}^0$ corresponds to the sum of contributions \eref{beta_g} for the original quiver, without extending it by global nodes. Notice that the ranks that enter in $\beta_{tot}^0$ are not only the result of the regular D3 and wrapped D5-branes but are also modified if flavor D7-branes are included. In the second line, $X_i$ is the mesonic field that couples to fundamental flavors for a given $D7$-brane, and we have used that the R-charge of the corresponding superpotential term is equal to $2$.\footnote{Once conformal invariance is broken, there is no longer a map between dimensions and R-charges of operators. To keep expressions short, we have chosen to express \eref{beta_extended} in terms of R-charges rather than anomalous dimensions. The reader should be cautious and keep in mind that what we really mean is that we should write \eref{beta_extended} in terms of anomalous dimensions, derive the result \eref{beta_D7} concluding that corrections to anomalous dimensions do not enter $\beta_{tot}$ to leading order and then switch to the more compact notation in terms of $R$.}

So far, the only thing we know is that $\beta_{tot}$ is zero in the absence of flavor D7-branes, with the first non-trivial contribution potentially appearing at ${\cal O}(K)$. Both terms in \eref{beta_extended} can a priori contribute. Since, as we have already mentioned, the ranks of gauge groups are modified in the presence of flavor D7-branes (in addition, this also changes the number of flavors each gauge group sees coming from bifundamentals connecting to other nodes) it is a in principle possible that $\beta_{tot}^0$ provides a non-vanishing contribution to the beta function.

Let us compute the linear ${\cal O}(K)$ term in $\beta_{tot}$. To calculate the contribution of $\beta_{tot}^0$ to this linear term, we can find the beta function for each gauge node and add them up.\footnote{Notice that while we focus on the linear term of the beta function, we are {\it not} assuming that either $\gamma^{(1,0})$ or $\gamma^{(0,1)}$ vanish.} We now introduce the additional input that the gauge theory we are considering is not an arbitrary quiver, but it is engineered with D-branes on a singularity. As explained, each node in the quiver corresponds to a bound state of D3, D5 and D7-branes (with the latter wrapped over compact 2 and 4-cycles, respectively). Since the D7-brane components do not extend radially, they do not contribute to a radial dependence of the dilaton. We conclude that $\beta_{tot}^0=0$. We are left with the simple expression


\beq
\beta_{tot}=-{3 \over 2}\sum_{D7_i}R_{X_i} \, .
\label{beta_D7}
\eeq
An interesting consequence of this result is that, even in the case $\gamma^{(1,0)}$ and/or $\gamma^{(0,1)}$ do not vanish, they do not contribute to $\beta_{tot}$ to linear order. In section 6 we rederive this result from gravity.


\subsection{Ignition scale}

We now have all the tools necessary for estimating the decoupling scale $\Lambda_c$. As discussed in section \ref{section_introduction}, it is sufficient to follow $x_{tot}$, avoiding the intricate evolution of independent gauge couplings along the cascade.\footnote{$x_{tot}$ only involves the non-abelian part of the gauge group. It is possible to trace $U(1)$ couplings along the entire RG-flow, but we do not consider them in our discussion.}

Let us denote IR and UV the energy regimes below and above $\Lambda_c$, respectively. In the IR, the theory is weakly coupled so it is sufficient to compute $\beta_{tot}^{IR}$ at 1-loop. On the other hand, the theory is strongly coupled above $\Lambda_c$ and $\beta_{tot}^{UV}$ is given by \eref{beta_D7}. 

The gauge theory below $\Lambda_c$ might contain additional charged fields beyond the matter content of the pure MSSM (e.g. four extra doublets in our example). These fields have to be heavier than roughly 1 TeV. Their effect on the RG-running is to reduce $\beta_{tot}^{IR}$ at energies above their masses, pushing the intersection scale $\Lambda_c$ to higher energies. As a result, it is sufficient to take $\beta_{tot}^{IR}$ to be given by the 
MSSM result in order to get a lower bound on $\Lambda_c$. This corresponds to taking the unknown new masses to be ${\cal O}(\Lambda_x)$. In other words, computing the matching using the IR running of the pure MSSM provides a lower bound for new physics, which can be either the appearance of new strongly coupled gauge groups or the more standard possibility of exotic heavy matter.

Let us discuss in some more detail the behavior of $x_{tot}$ around $\Lambda_c$. Following section \ref{section_deconstructed_dimension}, the G nodes confine and the ones in the two SM sets are pairwise higgsed to the diagonal SM. Since G is at infinite coupling, $x_{tot}$ only receives contributions from the two SM sets. Furthermore, each of the two SM copies contains a $U(1)$ node. We can think about them as containing trivial "SU(1)" factors with no associated gauge couplings. These nodes arise after dualizing $N_f=N_c+1$ gauge groups. These Seiberg dualities take place when their inverse couplings vanish. 
We conclude that, just above $\Lambda_c$, $x_{tot}$ only gets contributions from the two $SU(2)$ and two $SU(3)$ nodes in the SM copies. The inverse squared couplings of the diagonal groups are simply given by 
\beq
x_i^{(SM_D)} = x_i^{(SM_L)}+x_i^{(SM_R)}
\eeq
with $i=2,3$.

Let us proceed to the actual computation of $\Lambda_c$. The superconformal R-charges for SPP$/\IZ_3$ are

\beq
R_{\Phi}=2-{2\over \sqrt {3}} \ \ \ \ \ \ \ R_{Z,\tilde{Z}}=1-{1\over \sqrt {3}} \ \ \ \ \ \ \ R_{X,\tilde{X},Y,\tilde{Y}}={1\over \sqrt {3}} \, ,
\eeq
where we have used the notation of fields in the parent SPP theory. Plugging them into \eref{beta_D7} we obtain $\beta_{tot}^{UV}=-9$. Using the values of the $SU(2)$ and $SU(3)$ gauge couplings at the Z-pole \cite{pdg}, we get that the IR and UV runnings are given by

\beq
x_{tot}^{IR}=238.6+2\log\left(\Lambda / M_Z \right) \ \ \ \ \ \ \ \ \ x_{tot}^{UV} = -9 \log\left(\Lambda/\Lambda_w\right) \, .
\eeq
The only free parameter is the position of the duality wall $\Lambda_w$. Since the wall indicates the transition to the full 10-d string theory, it is natural for $\Lambda_w$ to sit close to $M_{Pl}$. For 
$\Lambda_w=10^{19}$ GeV we get $\Lambda_c\sim 3000$ TeV. We can contemplate lowering the wall to GUT scales. For $\Lambda_w=10^{16}$ GeV we get $\Lambda_c\sim 10$ TeV.
It is interesting to notice that the $\Lambda_c/\Lambda_w$ hierarchy is mostly determined by IR parameters: namely the D7-branes that are necessary to obtain a sensible MSS-like model. In addition, even though the decoupling scale can get interesting low values, it 
naturally comes out sufficiently high to avoid any compositeness constraint.

\section{Supergravity Dual}

We now turn to describe the properties of the supergravity
gravity dual of a cascading gauge theory. The specific example of interest is the
dual of the gauge theory associated with a collection of fractional  and flavor branes 
at the $\Zbb_3$ orbifold of the suspended pinch point singularity. The SPP singularity is the real cone over the  $L^{121}$ space, one of the members of the $L^{abc}$ series of 5-manifolds. The metric on this space is explicitly known \cite{Cvetic:2005ft,Martelli:2005wy}, and this in principle allows us to be rather explicit about the gravity dual of our specific example. We will try to keep our discussion somewhat general, however,
using the SPP/$\Zbb_3$ example as our guide.

The 6-dimensional internal manifold, on which the various branes will be wrapped, is an orbifold of a cone over 
an Einstein-Sasaki 5-manifold $\Sigma_5$, which we assume is an $L^{abc}$ space. The 5-manifold itself is a circle bundle over a 4-d base manifold $\Sigma_4$. 
The metric on the total 6-d cone takes the form
\begin{equation}
ds_6^2=dr^2  + r^2(d\psi +A)^2+r^2 ds^2_4\ ,
\end{equation}
where $ds_4^2 = h_{a\bar{b}}dz^a dz^{\bar{b}}$  is the metric of the 4-d 
K\"ahler-Einstein base manifold $\Sigma_4$,
with $R_{a\bar{b}} = 6 h_{a\bar{b}}$ and K\"ahler 2-form $J_{a\bar{b}} = 6 i R_{a\bar{b}}$.
The coordinate $\psi$ is the angular coordinate on the circle bundle of the base $\Sigma_4$ and
the one-form $A$ satisfies $$dA = 2 J.$$
For the non-compact cone, the $r$-coordinate has infinite range.
We assume that the
orbifold is defined via isometric identification map on the base manifold $\Sigma_5$.
The base orbifold $\Sigma_5/\Gamma$ contains various homology cycles, that may support 
wrapped D-branes and their associated RR-fluxes.

Our brane configuration involves fractional and flavor D7-branes.
The fractional branes  are bound states of D7, D5 and D3-branes that wrap compact cycles within the base $\Sigma_4$ of the cone,
whereas the flavor D7-branes are wrapping non-compact cycles. Correspondingly,
the world-volume gauge fields  of the fractional branes have normalizable zero modes 
and descend to local 4-d gauge fields, while for the
flavor branes, the zero mode of the world-volume gauge fields  is non-normalizable and
the 7-d gauge group descends to a global flavor symmetry in 4-dimensions.

\subsection{Gravity Dual without D7-branes}

To begin, let us summarize the asymptotic 
behavior of the solution dual to the cascading gauge theory associated with $N$ D3-branes  
and $M$ wrapped D5-branes, without any flavor D7-branes \cite{Herzog:2004tr,Martelli:2005wy,Gepner:2005zt,Sfetsos:2005kd}. As in the KS
solution, it is characterized by the presence of the imaginary anti-self-dual 3-form flux.

The 4-d  base manifold $\Sigma_4$ supports an anti-self-dual closed (1,1)-form $\omega_2$, 
satisfying $\star_4\, \omega_2=-\omega_2$. We normalize  $\omega_2$ to have unit period
around the dual 2-cycle ${\cal C}_2$.
It is then straightforward to show that the (2,1)-form
\be
\label{twoone}
\Omega_{2,1} = \kappa\, \bigl( \frac{dr}{r} + i (d\psi +  A)\bigr) \wedge \omega_2
\ee
is imaginary self-dual and primitive. Here $\kappa$ is a normalization constant, fixed such that $\Omega_{2,1}$ has unit period around its dual 3-cycle ${\cal C}_3$. With the help of this, we can now write the supersymmetric Ansatz for the
asymptotic metric and RR-field strength \cite{Giddings:2001yu}
\bea
\label{ktlike}
ds^2 \is \, \frac{1}{\sqrt{h}}\, dx_{1,3}^2\ +\, \sqrt{h}\; ds_6^2\ , \nonumber \\[2.5mm]
g_s F_5 \is\; (1+\star_{10})dh^{-1}\! \wedge d^4x\ .\\[3mm]
G_3\is i M\,  \Omega_{2,1}\, ,  \qquad \quad G_3 = dC_2 + \tau H_3 \nonumber
\eea
where $\tau = C_0 + i e^{-\phi}$ is the axio-dilaton, $F_p = dC_{p-1} - C_{p-3}\wedge H_3$ denote the RR p-form field strengths,
and $H_3 = dB$ is the field strength of the NS 2-form $B$. The solution for the 
warp factor $h$ follows from its relation with the self-dual 5-form field strength, which 
in turn is obtained by integrating its Bianchi identity 
\be
\label{dfive}
 dF_5 = H_3 \wedge F_3\, .
\ee
The shape of the warp factor dictates the holographic identification between the radial coordinate $r$ 
and the RG scale of the 4-dimensional gauge theory. The solution without D5-branes has a constant dilaton $e^\phi = g_s$ and axion $C_0$.

\newcommand{\Pfr}{\mbox{$\mathfrak{b}$}}

The above equations uniquely characterize the solution for large values of the radial coordinate $r$.
The supergravity solution can be trusted in the regime where the 3-form flux $M$ is large compared
to $1/g_s$.
Naive extrapolation to small r, however, yields a solution that develops a singularity at its tip.
This singularity needs to be
smoothed out via a suitable complex deformation, similar as for the KS solution. 
As explained in the introduction, however, for the specific application to the cascading
theory of interest, we are interested in the large radius region only, corresponding to regime II far from its IR starting point. In this region,
not only the supergravity approximation may be trusted but also the deformation is unimportant. 

A D5-brane wrapped around the 2-cycle ${\cal C}_2$ creates one unit of $F_3$ flux around
the dual 3-cycle ${\cal C}_3$. 
The solution without D7-branes has a fixed D5-brane charge $M$, read off by integrating
the RR 3-form field strength over the 3-cycle ${\cal C}_3$ within $\Sigma_5$ dual to  $\Omega_{2,1}$
\bea
M = \int_{{\cal C}_3} \! F_3 \, .
\eea
The D3-brane charge, however, is not fixed: it evolves as a function of the holographic RG-coordinate $r$. 
We can introduce the effective D3-brane charge
as the RR 5-form flux through the 5-d base manifold $\Sigma_5$
\bea
N^{\rm eff} = \int_{\Sigma_5} F_5\, .
\eea
$F_5$ is sourced by the D3-branes, and as seen from eqn. (\ref{dfive}), by the
cross product of the 3-form fluxes. The $r$-dependence of $N^{\rm eff}$ can be expressed as
\be
\frac{d N^{\rm eff}}{dr}  \, = \, M\; \frac{d\Pfr}{dr} 
\ee
where 
\be
\Pfr = \int_{{\cal C}_2} B
\ee
 denotes the period of the NS 2-form around the 2-cycle ${\cal C}_2$.
 
Using the expression (\ref{ktlike}) for the imaginary anti-selfdual 3-form, we read off 
\be
\Pfr(r) \, \simeq  \, {\kappa\sp\sp g_s M}\sp \log r\, .
\ee
As shown in the original work of Klebanov and Strassler,
 the $r$-dependence  of the D3-brane charge precisely matches with the growth in the gauge group rank during the RG cascade of the dual gauge theory.  The period $\Pfr$ represents the difference between the inverse gauge couplings of 
 gauge groups, and its $r$ dependence reflects the running of these couplings. Evidently,
 the rate of growth in the total rank of the gauge groups during the RG cascade is proportional to the
 beta-functions that drives the gauge couplings.
The beta function is constant and the effective rank thus growth logarithmically
with scale.

\bigskip

\subsection{Adding the flavor D7-branes}

We now add the flavor D7-branes to this system. We will take into account their effect on the background to linearized order, in very much the same spirit of \cite{Ouyang:2003df}. This is an allowed approximation in the supergravity regime II, where the D5-brane
charge $M$ is much larger than $K$, the number of D7-branes. 

In Einstein frame, the D7-brane world volume action reads
\bea
\label{world}
 \int_{D7} e^\phi \, \sqrt{-\det(\hat g+ e^{-\phi/2} \, \calF)} \; + \int_{D7} \hat C_q \, e^{-\calF}
\eea 
It depends on the gauge field strength $F$ via the
combination 
\be
{\cal F} = F - B\, .
\ee
Flavor D7-branes are space-time filling. The supersymmetric branes wrap a holomorphic, non-compact 4-cycle within the 6-manifold, specified
by some embedding equation
\bea
\alpha(z_i) = 0 \, ,
\eea
with $\alpha(z_i)$ some holomorphic function defined on the internal 6-manifold.  Although we will not attempt to do so here,
it should in principle be possible to find the
explicit form of the necessary embedding equations
for the specific example of the SPP/$\Zbb_3$ quiver discussed in this paper. We will assume that 
$\alpha(z_i)$ is single-valued.

Since the D7-branes are a magnetic source for the axion field strength $F_1 = dC_0$,
the axio-dilaton has non-trivial monodromy around the D7-brane locations. A holomorphic
expression for the axio-dilaton with the proper monodromy is \cite{Grana:2001xn,Burrington:2004id}
\be
\label{axiodil}
\tau = \frac{i}{g_s} - \frac{i }{2\pi } \log \alpha(z_i).
\ee
Notice that when obtaining \eref{axiodil} we are just imposing that $\tau$ has the correct $SL(2,Z)$ monodromy. We therefore expect this expression to be valid even beyond the linearized regime. Indeed, this is the case in known flavored examples where the full backreacted solution has been obtained \cite{Benini:2006hh,Benini:2007gx,Benini:2007kg}. If $S^1$ denotes a circle that surrounds $K$ D7-branes, that is, if the function $\alpha(z_i)$ has
$K$ zeroes within the region surrounded by the $S^1$, then
\bea
\oint d\tau = 
\oint_{S^1} F_1 = K
\eea

The holomorphic function $\alpha(z_i)$ can be factorized into a product of $K$ elementary functions $\alpha_k(z_i)$, each representing the embedding equation of a single elementary D7-brane.
As previously explained, these embedding equations arise via the F-term equations
$X_k=0$ of the extra flavored matter associated with the intersection between the flavor and
the fractional branes. This correspondence motivates the interpretation of the $\alpha_k(z_i)$
as the holographic wavefunctions of the operators $X_k$ associated with the corresponding flavor node \cite{Berenstein:2005xa,Oota:2005mr}.
We can thus relate the asymptotic dependence of the embedding function $\alpha(z_i)$, as a function
of the radial coordinate $r$ and angular coordinate $\psi$, to the sum of the scaling dimensions
and R-charges of the operators $X_k$,
via
\be
\label{alphaz}
\alpha(z_i) \sp = \, z_1^{\beta} \, \tilde{ \alpha}(x_i) 
\ee
where $z_1 = r e^{i\psi/3}$ and
\be
\beta = - \frac{3}{2} \sum_{D7_k} R_{X_k}
\ee
Plugging this into the expression for the dilaton gives
\be
\label{dilatonRG}
e^{-\phi} = \frac{1 }{g_s} -\frac{\beta}{2\pi } \sp \log(r/r_0)
\ee
Here we recognize the RG dependence of the total gauge coupling $x_{\rm tot} = \sum_k x_k$
of the quiver gauge theory found in the previous section. We will explain this correspondence  in 
the next subsection.

The combined presence of fractional  and D7-branes leads to some subtle modification in the equations for the higher RR-forms $F_3$ and $F_5$. Since the NS 3-form flux is turned on, its pull-back to the D7 worldvolume will be non-zero. This will in turn induce worldvolume flux ${\cal F}$, that sources the RR-forms. Taking this
into account, one derives the following Bianchi identitites
\bea
dF_1 \is -  \Omega_2 
\nonumber \\[2mm]
dF_3 \is H_3 \wedge F_1 - \calF\wedge\Omega_2 \
\\[2mm]
dF_5 \is H_3 \wedge F_3 - \frac{1}{2} \calF\wedge\calF\wedge\Omega_2 \nonumber
\eea
Here $\Omega_2$ denotes the delta-function 2-form localized on
the D7-brane world-volume. The extra localized terms reflect the fact that the flux of $\calF$
and $\halff \calF \wedge \calF$ give rise to extra induced D5-charge and D3-charge.
The presence of the localized source terms can also be understood by requiring that the $SL(2,\Zbb)$ covariant 3-form field $G_3 = F_3 + \tau H_3$ remains single-valued around the D7-branes.

More important for our purpose, however, is that the above equations show that, in the presence of
D7-branes,
the flux of the RR 3-form $F_3$ is no longer quantized or constant.
We can introduce an effective D5-brane charge 
\be
M^{\rm eff} = \int_{{\cal C}_3} F_3
\ee
Here the integral is performed at a given radial location $r$. Let us suppose that the $F_1$ flux through 
the $S^1$ at this location $r$ take the quantized value $K$. It is then straightforward to derive that
the effective brane charges can be expressed in terms of the period $\Pfr$ of the NS B-field as
\bea
\label{effcharge}
M^{\rm eff} \is M + K \Pfr \nonumber \\[-2mm]
&& \\[-2mm]
N^{\rm eff} \is N + M \Pfr + \halff K \Pfr^2 \nonumber
\eea
The radial dependence of the period $\Pfr$ can be found by integrating the Bianchi equation
for the $SL(2,\Zbb)$ covariant 3-form
\be
d G_3 = d \tau \wedge H_3\, .
\ee
In the linearized approximation, and using eqns. (\ref{twoone}), (\ref{ktlike}) and (\ref{axiodil}), this gives
\be
d\sp G_3 \, \simeq \, \frac{  \kappa\sp\sp g_s M}{2\pi i} \, \frac{d\alpha}{\alpha} \wedge \frac{dr}{r} \wedge \omega_2
\ee
 This equation may be integrated, with the result
\be
\label{bflow}
\Pfr(r)\, \simeq \, \kappa \sp\sp g_s M\sp \bigl(  \log r \,  + \frac{\beta }{2\pi}\, (\log r)^2 \sp + \ldots\, \bigr)
\ee
where we used eqn. (\ref{alphaz}).

The equations (\ref{bflow}) and (\ref{effcharge}) summarize the holographic 
manifestation of the RG cascade of the dual quiver gauge theory,
and show how the presence of the extra flavors accelerates the RG running and
of the growth in the number of colors. We will now elaborate this correspondence,
and make a more detailed comparison with the gauge theory description of the duality
steps. 

\bigskip

\subsection{Geometric description of the Duality Cascade}

We wish to recover the properties of the duality cascade of the quiver gauge theory from the string dual. To this end, we will make use
of two different geometrical perspectives:  

\smallskip
\medskip

\noindent
${}\, $  \parbox{15.5cm}{ \addtolength{\baselineskip}{.4mm}
(i) {\it D-brane-probe perspective.} The cascading gauge theory has not yet reached the large $N$
regime. It can be engineered from fractional and flavor  D7-branes on a singular cone geometry. Coupling constants are parametrized by the geometric data defined on the cone. RG-running is represented by adjusting these 
geometric data.}

\medskip
\smallskip

\noindent
${}\, $ \  \parbox{15.5cm}{ \addtolength{\baselineskip}{.4mm}
(ii) {\it Supergravity perspective.} The cascading gauge theory is assumed to be in the large $N$ and large 't Hooft coupling regime. All degrees of freedom and couplings of the gauge theory 
can be captured by a dual geometry, with a number of flavor D7-branes. The fractional
brane charge is carried by RR-flux, as summarized in section~6.2.  }

\medskip
\smallskip

\noindent
The supergravity dual correctly encodes global properties of the cascading 
quiver gauge theory, such as the RG running of couplings and overall growth in the number of
colors, but does not appear to have sufficient structure to fully reconstruct
the sequence of Seiberg duality maps. The D-brane probe picture, on the other hand,
provides an attractive dual perspective on Seiberg duality, and thereby gives a useful 
interpolation between the gauge theory and supergravity description of the cascade.

\smallskip

How does one recover the duality cascade from the D-brane probe perspective?
The fractional branes are bound states of D3, D5 and D7-branes that wrap compact cycles.
within the 4-d base $\Sigma_4$ of 
the complex cone. The fluxes of $F$ along the non-trivial 2-cycles 
in the base represent
the D5-brane wrapping charges of the fractional brane; the D3-brane charge is the instanton number.
The D7, D5 and D3 wrapping numbers are combined in a charge vector, which we denote
by 
\be
{\cal Q} = (Q_3, Q_5,Q_7)
\ee
The number of independent fractional branes is equal to
the total dimension of the cohomology groups of the base $\Sigma$.
The quiver gauge theory is engineered from a collection of several intersecting fractional and flavor branes,
with total charge 
\be
\label{qtot}
\sum_{i} N_i \, {\cal Q}_i \, = \, (N, M, K \sp ) 
\ee
Here $N_i$ denotes the multiplicity of the  fractional brane with charge vector $Q_i$, and equals
the number of colors of the corresponding gauge group factor.

The gauge coupling of a fractional brane with charge vector $Q_i$ is given by \cite{Buican:2006sn}
\bea
\label{zcoupling}
x_i =  e^{-\phi} |Z({\cal Q}_{i})| 
\eea
where $Z(Q_i)$ is the central charge of the brane. The central charge is a linear functional
of the charge vector
\bea
\label{linear}
Z({\cal Q})= 
Q_3+ Q_5\Pi + Q_7\, {\cal K} . 
\eea
The formula (\ref{zcoupling}) can be derived from the boundary CFT description of fractional
branes, and does not rely on the supergravity approximation.
In  the geometric regime, the explicit form of the central charge can be found by expanding the
Born-Infeld world-brane action (\ref{world}).  One finds that $\Pi$ and ${\cal K}$ are  given by
the period integrals
\bea
\Pi \is \int_{{\cal C}_2}  (B+ iJ)\, , \nonumber \\[2mm] 
{\cal K}  \is  \frac{1}{2}\int_{\Sigma_4} (B + iJ)^2\, ,
\eea
where $J$ denotes the K\"ahler 2-form of the 4-d base manifold $\Sigma_4$. The holographic RG running of the gauge couplings 
is driven by the evolution of these integrals and the dilaton along the radial direction. 
Note that since the flavor D7-branes wrap a non-compact 4-cycle, the volume integral ${\cal K}$
diverges and the corresponding gauge couplings vanish.

The central charge $Z(Q)$  specifies which ${\cal N}=1$ sub-algebra of the ${\cal N} =2$
supersymmetry of the type IIA supergravity background is preserved by the fractional brane. A collection
of  branes is supersymmetric if all their central charges are aligned and have the same
sign.\footnote{A relative phase between central charges of two fractional branes 
translates into an FI-parameter of the gauge theory. Misalignment of the charges reflects supersymmetry breaking via D-terms.} This supersymmetry condition, that the central charges remain aligned throughout the holographic 
RG flow, naturally dictates the appearance of a duality cascade. As the periods
$\Pi$ and ${\cal K}$ evolve along the radial direction, the central charge vector of 
one or more fractional
branes may become zero at some value of the radial coordinate and change sign. Beyond this point,
the original configuration of fractional branes becomes unstable, and has to rearrange itself
into a stable configuration with aligned central charges. This rearrangement of the basis
of fractional branes is the geometric manifestation of the Seiberg duality map. 

Consider the location where the central charge of the fractional brane with charge vector $Q_j$ 
changes sign. Roughly speaking, it then turns into its own anti-brane, and its unbroken
supersymmetry becomes incompatible with that of the other fractional branes.
To restore supersymmetry, and stability, the fractional branes then form a new set of bound states.
  After this rearrangement 
charge vector of the $Q_j$ fractional brane has changed sign
\be
\label{qflip}
Q_j \to - Q_j
\ee
and, as suggested by the Seiberg duality map, has a new positive multiplicity equal to
\beq
\label{nflip}
N_j \; \to\;  F_j - N_j 
\eeq
where  $F_j$ denotes total number of flavors of the $Q_j$ node. We can express $N_j$ as
\beq
\label{nflavor}
F_j\, = \sum_{i>j} \, N_i \, f_{ij}
\eeq
where the sum runs over all other fractional branes that connect to the $Q_j$ fractional brane via a positive number $f_{ij}$ of incoming lines. To absorb the overall change in the charge vector,
these other branes form an appropriate bound state with $Q_j$ branes, 
adjusting their charge vector according to\footnote{The index $j$ is not summed over. 
Note further that the bound state formation changes
the expression (\ref{zcoupling}) for the gauge couplings of the nodes for $i>j$. The 
{\it value}
of the coupling, however, is continuous, since the transition 
(\ref{nbound}) occurs at the specific radial location where $Z(Q_j) =0$.}
\beq
\label{nbound}
Q_i^A \to Q_i^A - f_{ij}\, Q_j^A\, . 
\eeq
All fractional branes except the $Q_j$ brane do not change their multiplicity. From the geometric rules for deriving the chiral matter content, one can show 
that the above re-combination of bound states  reproduces the full Seiberg
duality map. Similar discussions can be found in \cite{Cachazo:2001sg,Feng:2002kk}.

It is important to note that this readjustment of fractional branes is not just some similarity transformation,
but a dynamical process. It realigns the central charges, ensuring
that supersymmetry and stability is maintained.\footnote{In the mathematical literature, the stable basis of
fractional branes is called an `exceptional collection', and the Seiberg duality map is a special
map known as a `mutation' \cite{Herzog:2003zc,Herzog:2004qw}.}
The new charge vectors and multiplicities still satisfy 
the constraint (\ref{qtot}): the total brane charge is preserved.

At first sight, this seems surprising.
After all, after many Seiberg duality 
steps towards the UV, all the gauge group factors have acquired some large rank, and this should 
obviously translate into a large total D-brane charge. Moreover, in this regime, the gauge theory is expected to admit a dual supergravity description, which  is supported by a correspondingly large RR-flux. 

The  resolution is found by looking  the D7-brane worldvolume
action (\ref{world}). As seen from the Chern-Simons term, the D7-brane world-volume carries an 
{\it effective} D5 and D3-brane charge given by the flux of ${\cal F}$ and of $\halff \calF\wedge\calF$
through the corresponding cycles.  This effective charge, when summed over all fractional
and flavor branes, adds up to (\ref{effcharge}).  It grows along  with the period $\Pfr$ of the 
NS 2-form, and thus runs over a continuous range of values. 
We now explain how this continuous effective D-brane charge is related to the `true' quantized D-brane charge, given by the integral flux of $F$ and
of $\halff F\wedge F$. 

The D7 world-volume action is invariant under the redefinition $B\to  B + d \Lambda$, $A \to  A + \Lambda$, with $A$ the world-volme gauge field and $\Lambda$ any one-form. 
$\Lambda$ is allowed to multi-valued; the only restriction is that $d\Lambda$
belongs to an integral cohomology class.  Large gauge transformations of this type shift the 
period of $B$  and  the fluxes of $F$ by an integer. Assuming that they simultaneously act
on all fractional brane world-volumes, the complete transformation on the D-brane
charges reads
\bea
\label{intrafo}
\Pfr \, \to\,  \Pfr  \sp -\sp  n \qquad \qquad \ \ \ \, & \qquad \qquad \qquad & Q_{7}\, \to\,   Q_{{7}} \nonumber \\[3mm]
M \, \to\, M +  n\sp K \qquad \quad \ \ &
\qquad \qquad\qquad  & Q_{5} \, \to\;  Q_{5} \, +\, n\,  Q_{{7}} \\[2mm]
N\, \to \, N+ n\sp M  + \halff n^2 K & \qquad 
\qquad & Q_{3}\; \to\;  Q_{{3}} +  n\sp  Q_{5}  +\, \half n^2 \, Q_{7}\nonumber
\eea
We can use this  invariance to turn $\Pfr$ into an angular variable,
restricted to the interval between $0$ and $1$.

We can thus summarize the situation as follows.
The presence of $H_3$ flux implies that the period $\Pfr$ steadily 
grows along the holographic RG-direction. Via eqn. (\ref{zcoupling}), this evolution matches with 
the RG running of the difference between the inverse gauge couplings $x_i$ of the gauge 
group factors. In the process, two types of discrete 
transitions take place. Whenever one of the central charges $Z(Q_i)$ changes sign,
the fractional branes rearrange according to the Seiberg duality map (\ref{qflip}) -(\ref{nbound}).
Secondly, whenever $\Pfr$ becomes equal to 1, we may reset it to 0 by performing the integral
shift (\ref{intrafo}). The former transformation acts on the gauge theory data,
the multiplicities (= ranks of the gauge groups)
and intersection numbers (= amount of chiral matter), according to the Seiberg duality map;
the latter transformation does not act on the gauge theory data, but changes the total D-brane charge 
in a way that reflects the accumulative growth of the gauge
group ranks of the quiver gauge theory.

\subsection{Total gauge coupling and the dilaton}

Finally, let us discuss in some more detail why it is natural in this context to
consider the total gauge coupling and explain its relationship with the dilaton.
The argument is as follows. In region II, after the cascade has proceeded through 
several cycles upwards, the multiplicities $N_i$ of the fractional branes all become
large, and of the following form
\beq
N_i=N+r_i \, M +s_i \, K \, ,
\eeq
where $r_i$ and $s_i$ are constants that encode the modification of the ranks in the presence D5 and D7-branes.
From \eref{intrafo}, we see that the number D7-branes remains constant while the D5-branes 
grow linearly with the step of the cascade $n$ towards the UV. At the same time, the number of 
D3-branes grows quadratically with $n$. Then, we soon reach a regime in which $N \gg M,K$, and 
the effect of the fractional brane charge $M$ and D7-brane charge $K$ can be treated as small 
perturbations away from the conformal large $N$ theory of $N$ D3-branes. As a result, the central charge 
\eref{zcoupling} is dominated by the D3-branes. From the linear relation (\ref{zcoupling})-(\ref{linear}) 
between the gauge coupling and the charge vector of the associated brane, we see that  the total gauge coupling
\be
x_{\rm tot} = \sum_i x_i
\ee
in this quasi-conformal regime reduces with the help of eqn. (\ref{qtot}) to the anticipated result 
\be
\label{xtot}
x_{\rm tot} = e^{-\phi}.
\ee

\bigskip

\section{Conclusions}

We have outlined the general principles of N-ification, a new class of string 
(motivated) UV extension of the SM. N-ification models provide a precise identification 
of the new degrees of freedom corresponding 
to embedding the SM in a warped higher dimensional space and their interaction with the SM: they are 
holographically mapped to a growing tower of field theory degrees of freedom in a 
duality cascade. 

The N-ification program can also be viewed as the generation of a UV string theory 
starting from an IR gauge theory. As such, it can be applied to more general IR starting points, such as the MSSM 
augmented by some hidden sector. The complex deformation that terminates the cascade separates them in the extra dimensions. 
The two sectors are then coupled by higher dimension operators suppressed by the mass
scale of string states stretching between them. Interesting hidden sectors include SUSY breaking ones
as well as conformal field theories. The latter provide realizations of the unparticle scenario
in N-unified models \cite{Georgi:2007ek}. In fact, some of the models classified in \cite{Cascales:2005rj} are 
explicit examples of this idea. 

Flavor D7-branes are a usual ingredient in type IIB local constructions
of MSSM-like theories. As we have explained, they lead to an acceleration of duality cascades.
This is a welcome effect in a holographic realization of the 
SM, since it rapidly puts us in a large-N regime. The acceleration can be
understood from both a field theory and a gravity dual perspective by simply following
the evolution of the total gauge coupling and the dilaton, respectively. It leads
to a large hierarchy between $\Lambda_c$ (the scale at which the cascade is ignited) and 
a UV scale at which the theory transitions to the full 10-d string theory (which is 
naturally located close to the Planck scale). Interestingly, this hierarchy is determined
by IR data -- the number and type of D7-branes -- rather than allowing arbitrary values, associated
with some choice of fluxes, as in other constructions.\footnote{There also is some dependence on the extra gauge 
group $G$, which controls the extension of region I in Fig. 3.
This dependence vanishes in the limit of large rank for $G$.}

It would be interesting to investigate alternatives for SUSY breaking besides the possibility of a hidden sector. 
For example, whether the strong dynamics of the gauge group $G$ might be exploited for this purpose.

A natural question is whether an N-ification extension of the SM can not only accommodate but also constraint the 
SM gauge couplings to have the observed values that are alternatively interpreted as arising from a GUT at high energies. In this work we have only used very general properties of the duality cascade, such as the running of the total gauge coupling.
It is possible that the requirement of a periodic cascade imposes strict constraints on the allowed
couplings at $\Lambda_c$ . Not surprisingly, the simple analysis of 
\cite{Franco:2004jz} shows that in some cases periodic cascades rule out entire regions in the space of couplings.

\bigskip
\bigskip

\begin{center}
\bf{Acknowledgements}
\end{center}
\medskip

We would like to thank F. Benini, J. Heckman, C. Herzog, D. Martelli, G. Shiu, A. Uranga, C. Vafa and 
M. Wijnholt for useful discussions. S.F. is supported by the DOE under contract DE-FG02-91ER-40671. 
D. R-G. acknowledges financial support from the European Commission through Marie Curie OIF grant 
contract no. MOIF-CT-2006-38381. The work of H. V. is supported by the National Science Foundation 
under grant PHY-0243680. S. F. would like to thank Harvad University, the University of Texas at 
Austin and the Michigan Center for Theoretical Physics for hospitality during part of this project. D.R-G is grateful to the University of Oviedo for warm hospitality while
part of this work was done.


\newpage

\appendix

\section{D-branes on $\Cbb^3/\Zbb_3$} 

In this section we discuss massless SUSY D7-brane embeddings in $\IC^3/\IZ_3$. A similar discussion applies to flavor branes in other singularities.

\bigskip

\subsection*{D3-branes on $\Cbb^3/\Zbb_3$}  

The simplest example of a three node quiver theory is that
 of single D3-brane probe near a $\Cbb^3/\Zbb_3$ singularity. The orbifold identification
  gives rise to three `image branes',
that are interchanged under the $\Zbb_3$-action 
\be
(x_1,x_2,x_3) \to (\omega x_1,\omega x_2, \omega x_3)\, , \qquad \qquad
\omega^3=1.
\ee 
Open strings in the untwisted sector reconnect with the same brane, while the twisted sector
open strings connect two different image branes. In the limit where the D3-brane approaches the
orbifold fixed point, the three image branes
recombine into three so-called fractional branes, corresponding to the  three irreducible 
representations of the abelian orbifold group $\Zbb_3$. 

If we place $N$ D3-branes at the orbifold
singularity, each fractional brane will occur with  multiplicity $N$. The world-volume theory 
 takes the form of a quiver gauge theory with gauge group $U(N)^3$ and with three `generations' of bifundamental chiral matter, 
associated with the intersections between the fractional branes. The three matter generations are the inherited from the three complex matter fields that describe the motion
of a D3-brane on $\Cbb^3$.  The quiver diagram
is shown in Fig. 9, where we have indicated 
each generation with a different color.

The configuration space of the D3-brane, the $\Cbb^3/\Zbb_3$ geometry, is recovered
by considering the moduli space of the quiver gauge theory, given by the space of expectation values of the chiral
matter fields, modulo gauge transformations and F- and D-term equations. Invariant coordinates 
on $\Cbb^3/\Zbb_3$ are
\be
\label{zi}
z_r = \Tr(X^r X^r X^r)\, , \qquad \qquad \mbox{\footnotesize $r = 1,2,3$}.
\ee
Other gauge invariant monomials can be re-expressed in terms of the $z_i$, by
using trace identities and F-term equations. The superpotential of the orbifold theory is the one
inherited from 
the ${\cal N} =4$ SYM covering theory, 
$W = \epsilon_{ijk} {\Tr}(X^i X^j X^k).$

\begin{figure}[t]
\begin{center}
 \epsfig{figure=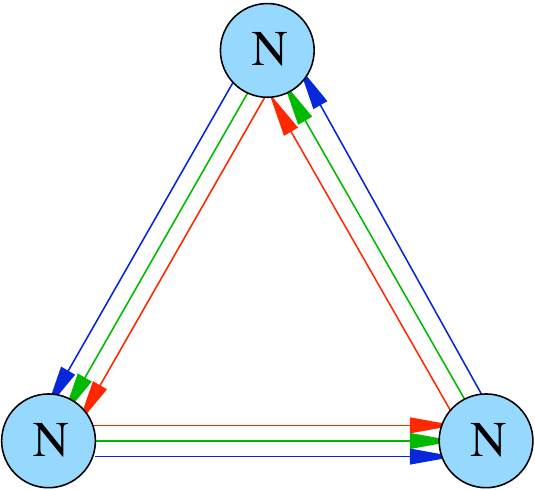,scale=0.9}
\end{center}
  \caption{The quiver diagram of $N$ D3-branes on a  $\Cbb^3/\Zbb_3$ orbifold.} 
\end{figure}

In D-brane engineering of quiver gauge theories,
the individual fractional branes can be used as independent building blocks. By 
assigning each fractional brane its own multiplicity $n_i$,  one obtains
a quiver theory with gauge group 
$U(N_1) \times U(N_2) \times U(N_3)$. To get the Standard Model gauge group, we
introduce six  fractional branes, and let the $\Zbb_3$ orbifold symmetry act
via the Chan-Paton matrices 
\bea
\label{chp}
\gamma_{{}_{D3}} & = & {\rm diag} (\onebb_1,\omega \onebb_2,\omega^2 \onebb_3).
\eea
In other words: one fractional brane is in the identity representation $\onebb$ of $\Zbb_3$,
two branes  are in the $\omega \onebb$ representation, and three branes are in the $\omega^2 \onebb$ representation.
This choice is invariant under $U(1) \times U(2) \times U(3)$ gauge symmetry.
When done in isolation, the particular choice (\ref{chp}) of Chan-Paton matrices would be inconsistent:
the multiplicities $N_i$  must satisfy the constraint that the resulting gauge theory is free on non-abelian anomalies.
To arrive at a self-consistent construction, it is necessary introduce
additional branes that, via their intersections with the fractional D3-branes, 
give rise to the other chiral matter fields that contribute to anomaly cancelation in the Standard Model.
These extra branes are the flavor D7-branes. From a string theory point of view, the flavor D7-branes are required in order to cancel twisted RR tadpoles.

\medskip

\subsection*{D7-branes in $\IC^3/\IZ_3$}

D7-branes span the 3+1-dimensional space-time and a 4-dimensional subspace of the internal target space.  Supersymmetry requires that this internal subspace is some holomorphic
4-cycle, specified by an embedding equation of the form 
\be
f(z_i)=0 
\ee 
with $f(z_i)$ some
holomorphic function defined on  $\Cbb^3/\Zbb_3$.
The simplest class of elementary D7-branes are
given by the embedding equations
\be
\label{emb}
z_r = X_{12}^r X_{23}^r X_{31}^r = 0\, .
\ee
This equation evidently factorizes into three separate equations
\be
\label{fact}
X^r_{12} = 0 \, \quad \cup \quad X^r_{23} = 0 \quad \cup \quad X^r_{31} = 0\, .
\ee
That is, eqn. (\ref{emb}) is implied, if one of the three eqns. (\ref{fact}) is satisfied. 

The flavor D7-branes intersect with the fractional branes and this produces
extra chiral matter fields $Q$, which arise as the ground states of the D3/D7 sector of open strings. 
The embedding equation (\ref{emb})  arises as a solution to 
the F-term equation obtained by varying the superpotential for the chiral fields $Q$.
 Each separate 
factor in (\ref{fact}) is associated  to a superpotential term
\beq
W= 
\tilde{Q}_{i}X^r_{i,i+1} Q_{i+1}\, .
\eeq
The factorized solution of the embedding equation indicates that the D7-brane in fact
divides into three components.  Geometrically, eqns. (\ref{fact}) are the condition 
that the D3-branes are located within one of these three components\footnote{A D3-brane  and D7-brane can
form a supersymmetric bound state, when the D3-brane is embedded inside the D7
worldvolume.}, so that the D3/D7 open 
strings have massless ground states. The three components of the D7-brane are distinguished via
the action of the $\Zbb_3$-transformations on the D7 Chan-Paton index. (the extra  chiral matter fields
transform under $\Zbb_3$ as $Q \to \omega Q, \tilde{Q} \to \omega \tilde{Q}$). This class of factorizable embedding is equivalent to the Ouyang embedding of D7-branes in the conifold \cite{Ouyang:2003df}.

Compound D7-branes are specified via an embedding equation of the form
\be
\label{cmpd}
X^r_{12} (X^{s_1}_{23} X^{t_1}_{31} + X^{s_2}_{23}X^{t_2}_{31}) = 0\, ,
\ee
with $s_1\neq s_2$ and $t_1\neq t_2$. Eqn. (\ref{cmpd}) factorizes as
\be
X^r_{12} = 0 \qquad \cup \qquad X^{s_1}_{23} X^{t_1}_{31} + X^{s_2}_{23}X^{t_2}_{31} = 0\, .
\ee
The second equation can not be factorized further, and arises as the F-term equation 
of a quartic superpotential term
\be
W =\tilde Q_2 (X^{s_1}_{23} X^{t_1}_{31} + X^{s_2}_{23}X^{t_2}_{31}) Q_{1}
\ee
We have set the coefficients of both terms in \eref{cmpd} to $1$. By tuning their ratio, we can make
the embedding (and consequently the gauge theory) arbitrarily close to one with a single term
in the superpotential. In that case, the second term becomes important only for very large vevs of the corresponding
fields. The non-factorizable embeddings are analogous to Kuperstein embedding in the conifold \cite{Kuperstein:2004hy}.

\begin{figure}[ht]
\begin{center}
 \epsfig{figure=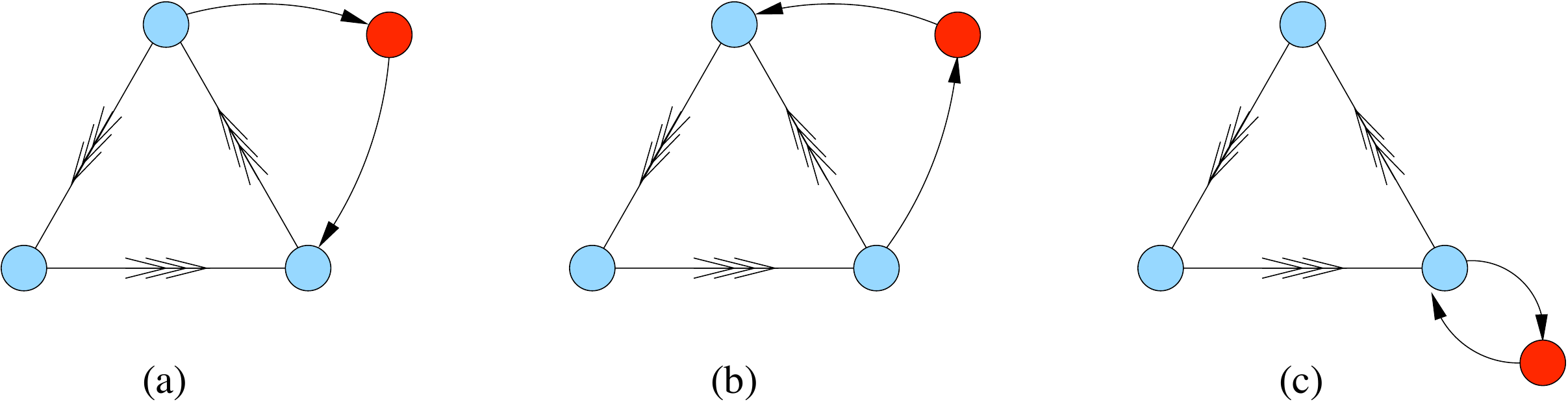,scale=0.5}
\end{center}
  \caption{The three classes of flavor D7-branes on $\IC^3/\IZ_3$. Notice the opposite orientation between (a) and (b). While class (a) couple to single bifundamentals, class (b) couple to linear combinations of products of two of them. Class (c) are completly non-chiral.}
  \label{D7s_C3_Z3}
\end{figure}

There are $27$ gauge invariant mesonic operators $a^{\alpha \beta \gamma}=X^{(\alpha)}_{12}X^{(\beta)}_{23}X^{(\gamma)}_{31}$ in this theory.\footnote{It is very useful to use the $SU(3)$ global symmetry of the theory to organize calculations \cite{Franco:2002mu}. The $X^{(\alpha)}_{ij}$ transform in the fundamental representation. The $a^{\alpha \beta \gamma}$ transform in $3\otimes 3 \otimes 3=1 \oplus 8 \oplus 8 \oplus 10$. Using the F-term equations, we conclude that any antisymmetric combination of the indices vanishes and only the $10$ survive. We are not going to exploit this clasification here.} Following the previous discussion, they can be used to define D7-branes of the three general classes shown in Fig. \ref{D7s_C3_Z3}. The D7-branes in Fig. \ref{D7s_C3_Z3}.c can be constructed with a straightforward extension of the above reasoning. They correspond to completely non-factorizable cubic embeddings. 

\newpage

\section{Geometry of $L^{abc}$} 

We are interested in 6-dimensional cones $\mathcal{C}L^{abc}$ over the $L^{abc}$ space \cite{Cvetic:2005ft,Martelli:2005wy}. Denoting by $ds_5^2$ the metric of the Sasaki-Einstein $L^{abc}$ manifold, we have

\begin{equation}
ds_5^2=(d\tau+A)^2+ds_4\ ,
\end{equation}
where $ds_4^2$ is the metric of the 4-dimensional K\"ahler-Einstein base. Explicitly, it reads

\begin{eqnarray}
ds_4^2=\frac{\rho^2}{4\Delta_x} dx^2+\frac{\rho^2}{\Delta_{\theta}} d\theta^2+\frac{\Delta_x}{\rho^2}\Big(\frac{\sin^2\theta}{\alpha} d\phi+\frac{\cos^2\theta}{\beta} d\psi\Big)^2+\nonumber\\ +\frac{\Delta_{\theta}\sin^2\theta\cos^2\theta}{\rho^2}\Big(\big(1-\frac{x}{\alpha}\big)d\phi-\big(1-\frac{x}{\beta}\big)d\psi\Big)^2\ ;
\end{eqnarray}
where 

\begin{equation}
\Delta_x=x(\alpha-x)(\beta-x)-\mu\ ; \quad \Delta_{\theta}=\alpha\cos^2\theta+\beta\sin^2\theta\ ;\quad\rho^2=\Delta_{\theta}-x\ .
\end{equation}
Here $0\le \theta\le \frac{\pi}{2}$, $x_1\le x\le x_2$, $0\le \psi,\phi\le2\pi$, being $x_1$, $x_2$, $x_3$ the roots of $\Delta_x$. We can introduce the parameters $a_i$, $b_i$, $c_i$, $i=1,2$ related to $\alpha$, $\beta$ as

\begin{equation}
a_i=\frac{\alpha c_i}{x_i-\alpha}\ , \quad b_i=\frac{\beta c_i}{x_i-\beta}\ , \quad c_i=\frac{(\alpha-x_i)(\beta-x_i)}{2(\alpha+\beta)x_i-\alpha\beta-3x_i^2}\ .
\end{equation}
Then $0\le\tau\le 2\pi|c_1|b^{-1}\,gcd(a,b)$. These $a_i, b_i, c_i$ are related to $a, b, c$ as

\begin{equation}
a\,a_1+b\,a_2+c=0\, \quad a\,b_1+b\,b_2+a+b-c=0\ ,\quad a\,c_1+b\,c_2=0\ .
\end{equation}
Also

\begin{equation}
\mu=x_1x_2x_3\ ,\quad \alpha+\beta=x_1+x_2+x_3\ ,\quad \alpha\beta=x_1x_2+x_1x_3+x_2x_3\ .
\end{equation}
Eventually, all parameters can be written in terms of the $a$, $b$ and $c$, giving rise to a nonsingular and complete manifold. In particular, the SPP$\sim L^{121}$ has

\begin{equation}
\alpha=\frac{3}{2}+\frac{\sqrt{3}}{2}\ ,\quad \beta=\sqrt{3}\ ,\quad x_1=-\frac{1}{2}+\frac{\sqrt{3}}{2}\ ,\quad x_2=1\ ,\quad x_3=1+\sqrt{3}\ .
\end{equation}

Our main example concerns the $\mathbb{Z}_3$ orbifold of the SPP singularity. In the local coordinates above, the orbifold amounts to certain global identification of the angular variables. Since we do not make use of any global property the of base manifold, for our purposes it is enough to consider the unorbifolded space. The results we obtain also hold in the orbifolded case.

The 1-form $A$ reads

\begin{equation}
A=\big(1-\frac{x}{\alpha}\big)\sin^2\theta d\phi+\big(1-\frac{x}{\beta}\big)\cos^2\theta d\psi\ ,
\end{equation}
and satisfies $dA=2J$, being $J$ the K\"ahler form of the K\"ahler-Einstein base. Defining the vielbein

\begin{eqnarray}
&&e^1=\frac{\rho}{\sqrt{\Delta_{\theta}}}d\theta\ ,\ e^2=\frac{\sqrt{\Delta_{\theta}}\sin\theta\cos\theta}{\rho}\Big(\big(1-\frac{x}{\alpha}\big) d\phi-\big(1-\frac{x}{\beta}\big)d\psi\Big)\ ,\nonumber \\ &&
e^3=\frac{\sqrt{\Delta_x}}{\rho}\Big(\frac{\sin^2\theta}{\alpha}d\phi+\frac{\cos^2\theta}{\beta}d\psi\Big)\ , \ e^4=\frac{\rho}{2\sqrt{\Delta_x}}dx\ , \ e^5=(d\tau+A)\ ,
\end{eqnarray}
the K\"ahler form reads simply $J=e^1\wedge e^2+e^3\wedge e^4$.

Let us introduce the following set of holomorphic closed 1-forms in the $L^{abc}$ \cite{Martelli:2005wy}

\begin{eqnarray}
\label{etas}
&&\eta^1=\frac{\alpha \cot\theta+\beta \tan\theta}{\Delta_{\theta}} d\theta+\frac{x(\alpha-\beta)}{2\Delta_x}dx+i(d\phi-d\psi)\ ;\nonumber \\
&&\eta^2=\frac{\alpha \cot\theta-\beta \tan\theta}{\Delta_{\theta}} d\theta+\frac{x(\alpha+\beta)-2\alpha\beta}{2\Delta_x}dx+i(d\phi-d\psi)\ ; \\
&&\eta^3=3\frac{dr}{r}+\frac{\alpha \cot\theta-\frac{3}{2}(\alpha-\beta) \sin2\theta-\beta \tan\theta}{\Delta_{\theta}} d\theta+\frac{3 x^2+\alpha\beta-2 x(\alpha+\beta)}{2\Delta_x}dx+id\psi_R\ .\nonumber
\end{eqnarray}
The angle $\psi_R$ is defined as $\psi_R=3\tau+\phi+\psi$, and it  is conjugated to the R-symmetry. This can be seen from the (3,0) form of the Calabi-Yau, which in this coordinates reads

\begin{equation}
\Omega=r^3\sin(2\theta)e^{i\psi_R}\sqrt{\Delta_x\Delta_{\theta}}\,\eta^1\wedge \eta^2\wedge \eta^3\ .
\end{equation}
Since the $\eta^i$ are closed, we can locally integrate them as $\eta^i=\frac{dz_i}{z_i}$. These $z_i$ define a set of complex coordinates on the Calabi-Yau. In fact, the holomorphic 3-form reduces to

\begin{equation}
\Omega=\frac{dz_1\wedge dz_2\wedge dz_3}{z_1z_2}\ .
\end{equation}

For later convenience, let us explicitly write the $z_i$ \cite{Martelli:2005wy}

\begin{equation}
\label{z}
z_1=\tan\theta f_1(x)e^{i(\phi-\psi)}\ ,\, z_2=\frac{\sin(2\theta)}{f_2(x)\Delta_{\theta}}e^{i(\phi+\psi)}\ ,\, z_3=r^3\sin(2\theta)\sqrt{\Delta_x\Delta_{\theta}}e^{i\psi_R}\ ;
\end{equation}
where the functions $f_1(x)$, $f_2(x)$ are defined as

\begin{equation}
f_1(x)=\Big(\exp\big(\int dx\frac{x}{2\Delta_x}\big)\Big)^{\alpha-\beta}\ ,\quad f_2(x)=\Big(\exp\big(\int dx\frac{1}{2\Delta_x}\big)\Big)^{2\alpha\beta}\Big(\exp\big(\int dx\frac{x}{2\Delta_x}\big)\Big)^{-(\alpha+\beta)}\ .
\end{equation}
Note that the radial coordinate, which will be dual to the energy scale in the field theory, only appears through $z_3$.

Various SUSY embeddings of D-brane probes in $AdS_5 \times L^{abc}$ were determined using $\kappa$-symmetry in \cite{Canoura:2006es}.


\end{document}